\documentclass[iop,apj]{emulateapj}

\usepackage{amsmath, amsthm, amssymb,amsfonts}
\usepackage[english]{babel}
\usepackage{natbib}
\usepackage{verbatim}
\usepackage{lipsum}
\usepackage{epstopdf}
\usepackage{graphicx}
\usepackage{soul}

\usepackage{xcolor}

\usepackage{hyperref}
\definecolor{darkblue}{rgb}{0.0,0.0,0.5}
\hypersetup{colorlinks,breaklinks,
            linkcolor=darkblue,urlcolor=darkblue,
            anchorcolor=darkblue,citecolor=darkblue}

\newcommand\bb[1]{\mbox{\boldmath{$#1$}}}
\newcommand\grad{\bb{\nabla}}
\newcommand\bcdot{\,\bb{\cdot}\,}
\newcommand\btimes{\,\bb{\times}\,}
\newcommand{\const}{{\rm const}}

\def\apjs{ApJS}
\def\apj{ApJ}
\def\apjl{ApJL}
\def\aap{A\&A}
\def\araa{ARA\&A}

\def\mnras{MNRAS}

\def\njp{NJPh}
\def\npgeo{NPGeo}

\def\pre{PhRvE}
\def\prl{PhRvL}
\def\prx{PhRvX}

\def\jgra{JGRA}

\def\frass{FrASS}

\def\pop{PhPl}

\def\pofa{PhFlA}

\def\jpp{JPlPh}
\def\jfm{JFM}

\begin{document}

\title{Turbulent regimes in collisions of 3D Alfv\'en-wave packets}

\author{S.~S.~Cerri$^1$}
\author{T.~Passot$^1$}
\author{D.~Laveder$^1$}
\author{P.-L.~Sulem$^1$}
\author{M.~W.~Kunz$^{2,3}$}
\affiliation{$^1$Université Côte d'Azur, Observatoire de la Côte d'Azur, CNRS, Laboratoire Lagrange, Bd de l'Observatoire, CS 34229, 06304 Nice cedex 4, France}
\affiliation{$^2$Department of Astrophysical Sciences, Princeton University, 4 Ivy Lane, Princeton, NJ 08544, USA}
\affiliation{$^3$Princeton Plasma Physics Laboratory, PO Box 451, Princeton, NJ 08543, USA}
\email{E-mail of corresponding author: silvio.cerri@oca.eu}

\begin{abstract}

Using three-dimensional gyro-fluid simulations, we revisit the problem of Alfv\'en-wave (AW) collisions as building blocks of the Alfv\'enic turbulent cascade and their interplay with magnetic reconnection at magnetohydrodynamic (MHD) scales. 
Depending on the large-scale value of the nonlinearity parameter $\chi_0$ (the ratio between AW linear propagation time and nonlinear turnover time), different regimes are observed.
For strong nonlinearities ($\chi_0\sim1$), turbulence is consistent with a dynamically aligned, critically balanced cascade---fluctuations exhibit a scale-dependent alignment $\sin\theta_{k_\perp}\propto k_\perp^{-1/4}$, resulting in a $k_\perp^{-3/2}$ spectrum and $k_\|\propto k_\perp^{1/2}$ spectral anisotropy.
At weaker nonlinearities (small $\chi_0$), a spectral break marking the transition between a large-scale weak regime and a small-scale $k_\perp^{-11/5}$ tearing-mediated range emerges, implying that dynamic alignment occurs also for weak nonlinearities.
At $\chi_0<1$ the alignment angle $\theta_{k_\perp}$ shows a stronger scale dependence than in the $\chi_0\sim1$ regime, namely $\sin\theta_{k_\perp}\propto k_\perp^{-1/2}$ at $\chi_0\sim0.5$, and $\sin\theta_{k_\perp}\propto k_\perp^{-1}$ at $\chi_0\sim0.1$. 
Dynamic alignment in the weak regime also modifies the large-scale spectrum, scaling approximately as $k_\perp^{-3/2}$ for $\chi_0\sim0.5$ and as $k_\perp^{-1}$ for $\chi_0\sim0.1$.
A phenomenological theory of dynamically aligned turbulence at weak nonlinearities that can explain these spectra and the transition to the tearing-mediated regime is provided; at small $\chi_0$, the strong scale dependence of the alignment angle combines with the increased lifetime of turbulent eddies to allow tearing to onset and mediate the cascade at scales that can be larger than those predicted for a critically balanced cascade by several orders of magnitude. Such a transition to tearing-mediated turbulence may even supplant the usual weak-to-strong transition.

\end{abstract}

\section{Introduction}\label{sec:Intro}

A wide range of space and astrophysical systems host turbulent plasmas~\citep[e.g.,][]{QuataertGruzinovAPJ1999,SchekochihinCowleyPOP2006,BrunoCarboneLRSP2013}.
The turbulent cascade transfers energy from the injection scales down to dissipation scales, where it is converted into heat and non-thermal particles, thus regulating the energetics and/or dynamics of a system. 
In the last decades, the properties of cascading fluctuations in weakly collisional plasmas have been explored in unprecedented detail thanks to {\em in-situ} measurements from spacecraft missions in the solar wind~\citep[e.g.,][]{GoldsteinARAA1995,AlexandrovaPRL2009,AlexandrovaPRE2021,PodestaJGR2009,SahraouiPRL2010,SahraouiRMPP2020,WicksMNRAS2010,WicksPRL2013,ChenJPP2016,BrunoCarboneLRSP2013,ChenAPJS2020,KasperPRL2021}.

At large (``fluid'') scales, the cascade may be described as MHD turbulence, with the building blocks of its Alfv\'enic component being interactions between counter-propagating Alfv\'en waves~\citep[e.g.,][]{IroshnikovAZ1963,KraichnanPOF1965,GoldreichSridharAPJ1995,Howes13,OughtonMatthaeusAPJ2020}. 
This Alfv\'enic cascade is naturally anisotropic with respect to the mean-magnetic-field direction, with field-parallel wavenumbers much less than their field-perpendicular counterparts, $k_\|\ll k_\perp$. Assuming a critical balance (CB) between the fluctuations' linear and nonlinear timescales, this cascade was originally predicted by \citet{GoldreichSridharAPJ1995} to exhibit a perpendicular spectrum ${\propto}k_\perp^{-5/3}$ and a spectral anisotropy $k_\|\propto k_\perp^{2/3}$, to which corresponds a parallel spectrum ${\propto}k_\|^{-2}$. 
Still within the CB assumption, the continuous shearing of fluctuations in the field-perpendicular plane associated with interactions between counter-propagating AW packets was later taken into account by \citet{Boldyrev2006}, postulating that fluctuations would be subject to a scale-dependent ``dynamic alignment'' (or anti-alignment) whose angle $\theta_k$ is such that $\sin\theta_k\propto k_\perp^{-1/4}$. This effect results in a 3D anisotropy of the turbulent fluctuations and a cascade whose spectrum follows $k_\perp^{-3/2}$, with a $k_\|\propto k_\perp^{1/2}$ spectral anisotropy (the $k_\|^{-2}$ spectrum being unaltered; in this case, $k_\perp$ is related to the shortest length-scale $\lambda$ of these 3D-anisotropic eddies, which is perpendicular to both the mean-field and magnetic-fluctuation direction; see \S\ref{sec:theory}).

Another fundamental aspect of plasma turbulence is the formation of current sheets (CSs), either as a result of large-scale, broad-band injection~\citep[e.g.,][]{PolitanoPOP1995,BiskampMuellerPOP2000,ZhdankinAPJ2013,SistiAA2021} or of direct AW-packet interactions~\citep[e.g.,][]{PezziJPP2017,VernieroJPP2018a,RipperdaJPP2021}. If sufficiently thin and long lived, these CSs can be disrupted by tearing and/or magnetic reconnection~\citep[e.g.,][]{CarbonePOF1990,ServidioNPG2011,ZhdankinAPJ2015,AgudeloRuedaJPP2021,RipperdaJPP2021}, processes that have been suggested to mediate the non-linear energy transfer at both MHD~\citep[e.g.,][]{CarbonePOF1990,BoldyrevLoureiroAPJ2017, MalletMNRAS2017,ComissoAPJ2018,DongPRL2018,TeneraniVelli2020} and kinetic~\citep[e.g.,][]{CerriCalifanoNJP2017,FranciAPJL2017,LoureiroBoldyrevAPJ2017,MalletJPP2017} scales. When this disruption occurs, we refer to the resulting turbulence as a ``tearing-mediated'' cascade, whose range at resistive-MHD scales is characterized by a steep $k_\perp^{-11/5}$ spectrum. The conditions under which a critically balanced, dynamically aligned cascade can mutate into a tearing-mediated cascade at a transition scale $\lambda_*\sim (k_\perp^*)^{-1}$ rely on two criteria: (i) that turbulent eddies are sheared in the field-perpendicular direction to set up a tearing-unstable configuration, and (ii) that these eddies live long enough to allow the tearing instability to grow and disrupt them. While the latter condition depends upon the material properties of the plasma (e.g., the resistivity $\eta$), the former is a consequence of the dynamic alignment of turbulent fluctuations that produces eddy anisotropy in the field-perpendicular plane.
In this context, regardless of whether dynamic alignment would have proceeded indefinitely until dissipation scales~\citep{PerezPRX2012} or would have been only a limited-range effect tied to the dynamics occurring at the outer scale~\citep{BeresnyakMNRAS2012}, what matters is that alignment occurs on enough scales to meet the condition for tearing instability to grow sufficiently fast (after which alignment  will anyway be -- partially or completely -- disrupted by reconnection events; see \S\ref{subsubsec:alignment_sim}).

In the fluid regime, the coexistence of a turbulent MHD inertial range with a steeper tearing-mediated regime at smaller (but still, fluid) scales has been evidenced within two-dimensional (2D) simulations~\citep{DongPRL2018}. In 3D, MHD simulations of the plasmoid instability in an inhomogeneous reconnection layer, leading to a self-sustained turbulent state, have however only reproduced the ``small-scale regime''~\citep{Huang2016}. 
A still-debated point concerns the existence of a tearing-mediated regime in 3D, where there are steep resolution requirements to separate clearly a so-called ``disruption scale'' (at which a tearing-mediated cascade would begin) from the actual dissipation scale when broad-band fluctuations are injected in the system. 
Here, we approach this problem by investigating interactions between counter-propagating AW packets. 
In this context, reduced models such as the two-field gyro-fluid (2fGF) model~\citep{PassotPOP2018,PassotSulemJPP2019,MiloshevichJPP2021,PSL_arxiv2022} can be extremely useful for isolating and modelling purely Alfv\'enic turbulence, without being affected by other modes and/or a plethora of kinetic effects~\citep[e.g.,][]{HowesPRL2011,ToldPRL2015,MatthaeusAPJL2016,CerriAPJL2017,CerriAPJL2018,CerriAPJ2021,GroseljAPJ2017,PerronePOP2018,ArzamasskiyAPJ2019,GonzalezPOP2019,SquireNatAs2022}.

In this work we provide the first evidence of a tearing-mediated cascade occurring at MHD scales due to the interaction of counter-propagating 3D AW packets. 
For weak initial nonlinearities, $\chi_0<1$, dynamic alignment of the relatively long-lived fluctuations leads to a strong, tearing-mediated cascade that replaces the more customary weak-to-strong turbulence transition. 
At $\chi_0\sim1$, a dynamically aligned, strong MHD turbulent regime establishes instead; a tearing-mediated cascade may eventually emerge, but not at the Lundquist numbers we are able to explore numerically. 
New scalings for weak turbulence subject to dynamic alignment and for the relevant transition scales are also provided.

\begin{figure*}[!ht]
\center%
\includegraphics[width=0.5\textwidth]{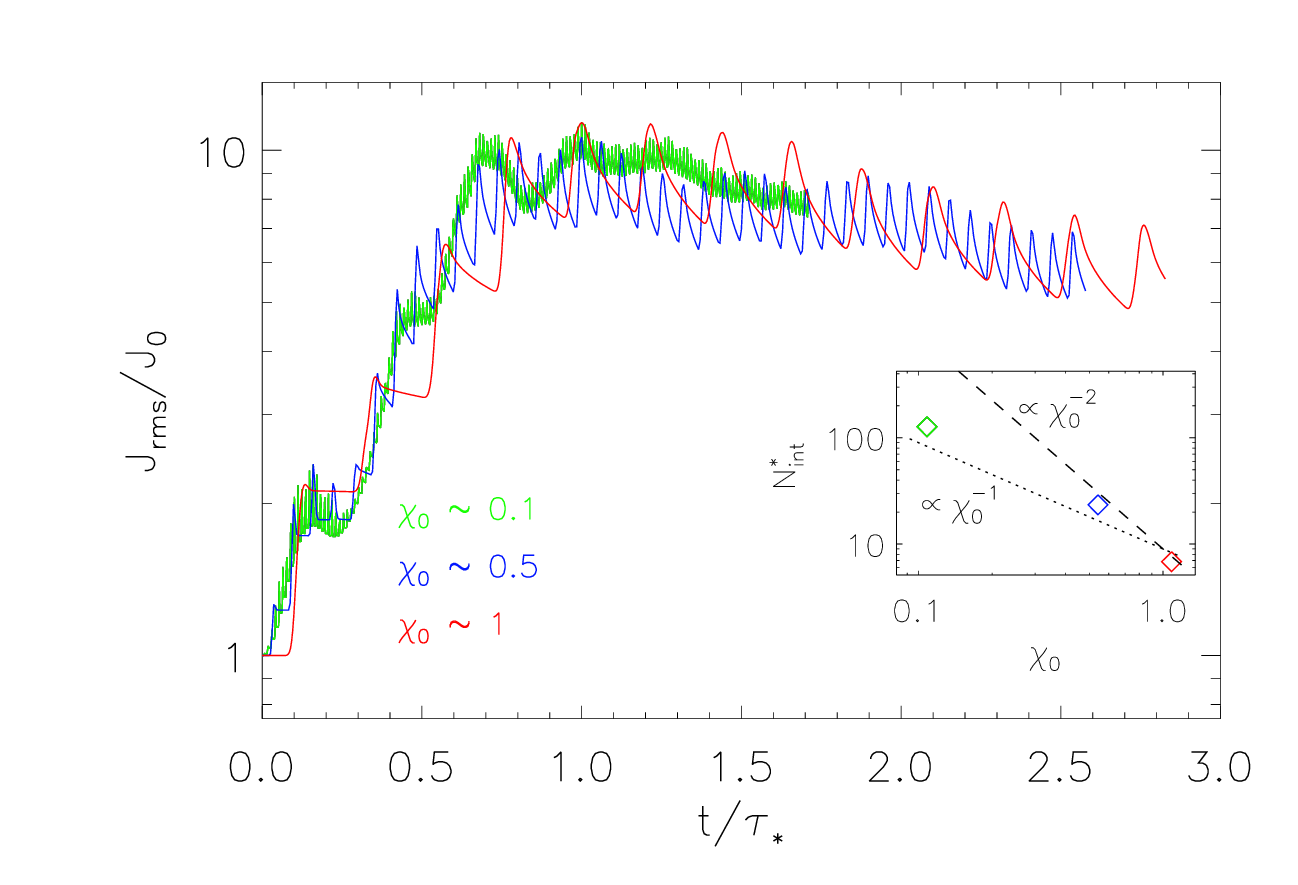}%
\includegraphics[width=0.5\textwidth]{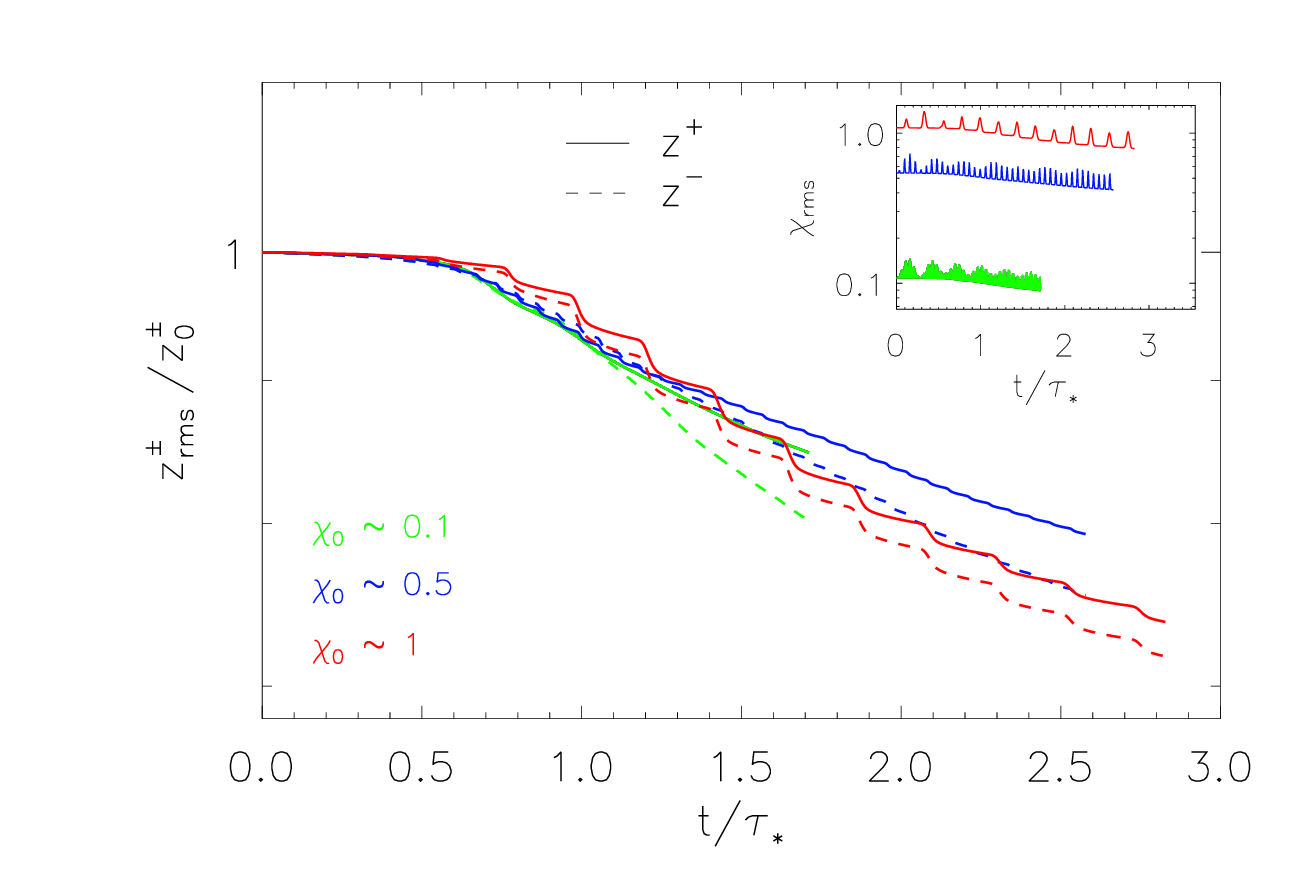}%
\caption{Left: time evolution of the root-mean-square (rms) current density, $J_{\rm rms}$ (normalized to $J_0=J_{\rm rms}(t=0)$; time is normalized to the time of peak activity $\tau_{*}$). The inset reports the (measured) number of collisions $N_{\rm int}^{*}=\tau_{*}/\tau_{\rm int}=3\tau_{*}/\tau_{\rm A}$ required to reach the peak of turbulent activity as a function of initial non-linearity parameter $\chi_0$, showing that a fully developed turbulent state is achieved on a timescale shorter than the one associated with the usual weak scaling (using the estimate $\tau_*\sim3\,\tau_{\rm nl}$, dotted line is $9/\chi_0$; a dashed line $9/\chi_0^2$ based on the estimate $\tau_*\sim3\,\tau_{\rm casc}$ is also given for reference). Right: time evolution of the rms Els\"asser fields, $z_{\rm rms}^{\pm}$ (normalized to $z_0^\pm=z_{\rm rms}^\pm(t=0)$). The inset shows the time evolution of the rms non-linearity parameter $\chi_{\rm rms}$.}
\label{fig:Jrms_compare_hst}  
\end{figure*}

\section{Two-field gyro-fluid simulations}\label{sec:GFmodel}

\subsection{Model equations}

To investigate nonlinear interactions between AW packets and the resulting multi-scale turbulent cascade, we employ the two-field gyro-fluid (2fGF) model~\citep{PassotPOP2018}, in which small-amplitude, low-frequency fluctuations are taken to be spatially anisotropic with respect to a mean magnetic field ({\em viz.}~$k_\perp\gg k_\|$, where $k_\perp$ and $k_\|$ are the wavenumbers perpendicular and parallel to the mean field, respectively). Although this model in general includes the finite inertia of the electrons, here we consider scales such that $k_\perp d_{\rm e}\ll 1$, where $d_{\rm e}$ is the electron skin depth. Finite electron-inertia effects can then be neglected and the equations for the number density of electron gyro-centers, $N_{\rm e}$, and the field-parallel component of magnetic potential, $A_\|$, read
\begin{equation}\label{eq:GF_Ne_delta0}
 \frac{\partial\,N_{\rm e}}{\partial t}\,
 +\,\left[\varphi,N_{\rm e}\right]\,
 -\,\left[B_z,N_{\rm e}\right]\,
 +\,\frac{2}{\beta_{\rm e}}\nabla_\|\Delta_\perp A_\|\,
 =\,0\,,
\end{equation}
\begin{equation}\label{eq:GF_Apara_delta0}
 \frac{\partial\,A_\|}{\partial t}\,
 +\,\nabla_\|\left(\varphi-N_{\rm e}-B_z\right)\,
 =\,0\,,
\end{equation}
where the Poisson bracket of two fields $F$ and $G$ is defined as $[F,G]\doteq(\partial_xF)(\partial_yG)-(\partial_yF)(\partial_xG)$, $\Delta_\perp\doteq\partial_{xx}+\partial_{yy}$ is the Laplacian operator acting perpendicular to $\bb{B}_0\doteq B_0\bb{e}_z$, and the electrostatic potential $\varphi$ and parallel magnetic-field fluctuations $B_z$ are related by $B_z=M_1\varphi$ and $N_{\rm e}=-M_2\varphi$.
The operators $M_1$ and $M_2$ are represented in Fourier space by ${\widehat M}_1 \doteq L_1^{-1}L_2$ and ${\widehat M}_2 \doteq L_3 + L_4L_1^{-1}L_2$, where $L_1\doteq2/\beta_{\rm e}+(1+2\tau)(\Gamma_0-\Gamma_1)$, $L_2\doteq1+(1-\Gamma_0)/\tau-\Gamma_0+\Gamma_1$, $L_3\doteq(1-\Gamma_0)/\tau$, and $L_4\doteq1-\Gamma_0+\Gamma_1$.
Here, $\beta_{\rm e}=8\pi n_0T_{{\rm e}0}/B_0^2$ is the electron plasma beta, $\tau=T_{{\rm i}0}/T_{{\rm e}0}$ is the ion-to-electron temperature ratio (so $\beta_{\rm i}=\tau \beta_{\rm e}$), and $\Gamma_n(b)\doteq {\rm I}_n(b)\exp(-b)$, with ${\rm I}_n$ being the first-type modified Bessel function of order $n$ and argument $b\doteq k_\perp^2\rho_{\rm i}^2/2$ ($\rho_{\rm i}\doteq v_{\rm th,i}/\Omega_{{\rm i}0}$ is the ion gyro-radius and $v_{\rm th,i}\doteq\sqrt{2T_{{\rm i}0}/m_{\rm i}}$ the ion thermal speed). 
Equations~\eqref{eq:GF_Ne_delta0} and \eqref{eq:GF_Apara_delta0} are normalized in terms of the ion-cyclotron frequency $\Omega_{{\rm i}0}\doteq eB_0/m_{\rm i}c$ and the ion-sound gyro-radius $\rho_{\rm s}\doteq c_{\rm s}/\Omega_{{\rm i}0}$, where $c_{\rm s}\doteq\sqrt{T_{{\rm e}0}/m_{\rm i}}$ is the ion-sound speed.

The 2fGF model effectively reproduces the so-called reduced magnetohydrodynamics (RMHD), or Hall reduced magnetohydrodynamics (HRMHD) if $\tau\ll1$, when employed at perpendicular scales much larger than the ion gyro-radius, $k_\perp\rho_{\rm i}\ll1$~\citep[see][for various limits of the 2fGF model]{PassotSulemJPP2019}. 
Our choice to employ the 2fGF model at MHD scales is motivated by the fact that it allows us to extend our investigations self-consistently to include interactions between Alfv\'enic wave packets at and below ion scales, which is the subject of a forthcoming paper.
The exact linear eigenmodes of the 2fGF system are given by the {\em generalised Els\"asser potentials}, $\mu^{\pm}\doteq\Lambda\varphi\,\pm\,\sqrt{2/\beta_{\rm e}}\,A_\|$, where $\Lambda\doteq (-\Delta_\perp)^{-1/2}(1+M_2-M_1)^{1/2}M_2^{1/2}$.
The associated {\em generalized Els\"asser fields} are $\bb{z}^\mp=\bb{e}_z\btimes\grad\mu^\pm$; they reduce to the usual \citet{ElsasserPR1950} fields in the MHD limit.

\subsection{Simulation setup}\label{subsec:sim_setup}

Equations (\ref{eq:GF_Ne_delta0}) and (\ref{eq:GF_Apara_delta0}) are discretized and solved on a $672^3$ grid for a $\beta_{\rm e}=\beta_{\rm i}=1$ plasma in a periodic cubic box of length $L_0=2\pi\rho_s\times\widetilde{\ell}_0$ with $\widetilde{\ell}_0=336$.\footnote{The numerical implementation of the 2fGF model adopts ``contracted variables'', i.e., quantities along the mean-field direction are re-scaled according to a gyro-fluid ordering parameter $\epsilon\ll1$. For example, $L_z^{\rm(code)}=\epsilon L_z^{(\rm real)}$ and  $k_z^{\rm(code)}= \epsilon^{-1} k_z^{(\rm real)}$. We have verified that an explicit choice of $\epsilon$ does not affect the following analysis.} 
A combination of second-order Laplacian dissipation (with resistivity $\eta$) and eighth-order hyper-dissipation operators removes energy close to the grid scale. 
Our choice of this operator combination and of their coefficients is such that (i) the dissipation scale is always above the ion scales, $k_{\rm diss}\rho_{\rm i}\lesssim1$, so that the inertial range of the cascade lies in the RMHD regime, and (ii) reconnection, at least at small values of $\chi_0<1$, is driven by the Laplacian resistivity $\eta$, i.e., there is enough range for a corresponding tearing-mediated cascade before achieving complete energy dissipation within the resolution thanks to the  hyper-resistivity.\footnote{This has been verified by running more than 50 simulations on a $560^3$ grid (keeping the resolution fixed by reducing $\widetilde{\ell}_0$) testing different combinations of dissipation operators and finding their optimal coefficients.See Appendix~\ref{app:numerical_tests} for a summary of these numerical tests.}
Nevertheless, we anticipate some differences at $\chi_0\sim1$ versus $\chi_0<1$, especially in that we do not expect to resolve $k_*$ (as predicted by \citet{LoureiroBoldyrevPRL2017} and \citet{MalletMNRAS2017}) when $\chi_0\sim 1$.

Two counter-propagating AW packets are initialized from the following potentials:
\begin{equation}\label{eq:initial_MUpm_counter}
 \mu^{\pm} =\,
  \mu_0^{\pm}\,\frac{\sin(\bb{k}_0^{\pm}\cdot\bb{x} + \psi^{\pm})}{|\bb{k}_{\perp,0}^{\pm}|}\exp\left[-\frac{1}{2}\left(\frac{z-\zeta_0^{\pm}}{\sigma_z^{\pm}}\right)^2\,\right]\,,
\end{equation}
where $\mu_0^{\pm}$ and $\bb{k}_0^\pm=k_{z,0}^\pm\bb{e}_z+\bb{k}_{\perp,0}^\pm$ are the initial amplitude and wavevector of the packets (centered at $z=\zeta_0^{\pm}$ with standard deviation $\sigma_z^{\pm}$), and $\psi^\pm$ is a random phase. The  packets' initial positions and widths are $\zeta_0^+=L_0/4$, $\zeta_0^-=3L_0/4$, and $\sigma_z^\pm=\widetilde{\ell}_0/3$. All simulations have the same initial amount of energy in the two Els\"asser fields, {\em viz.}~$\int |\bb{z}^+|^2 {\rm d}\bb{x}=\int |\bb{z}^-|^2 {\rm d}\bb{x}$, which is initially carried by modes $\bb{k}_0^{+}L_0/2\pi=(1,0,1)$ and $\bb{k}_0^{-}L_0/2\pi=(0,1,-2)$.
The slight asymmetry in $k_{z,0}^{\pm}$ causes a minor imbalance during the subsequent evolution (of order $\lesssim5\%$; Figure~\ref{fig:Jrms_compare_hst}, right panel). 
This is consistent with the von-Karman--Howarth decay law~\citep[e.g.,][and references therein]{WanJFM2012}, i.e., ${\rm d} (z^\pm)^2/{\rm d}t\propto-(z^\mp/\lambda_\pm)(z^\pm)^2$, with the similarity length estimated as $\lambda_\pm\sim 1/k_0^{\pm}$ (implying a slightly faster decay of $z^-$ in our setup).

Three different regimes defined by the initial non-linearity parameter of the AW packets are considered: $\chi_0\sim 0.1$, ${\sim}0.5$, and ${\sim}1$, where $\chi_0=\tau_{\rm lin}/\tau_{\rm nl}\approx(k_{\perp,0}\delta B_{\perp,0})/(k_{z,0}B_0)$~\citep[e.g., see][]{MiloshevichJPP2021}. The associated Lundquist numbers defined using the (second-order) resistivity $S_0=L_0v_{\rm A}/\eta$ are $\approx1.7\times10^6$, ${\approx}3.3\times10^5$, and ${\approx}1.7\times10^5$, respectively; these correspond to the same magnetic-Reynolds number $R_{\rm m}\doteq L_0 u_{\rm rms}/\eta\approx 2.2\times10^5$ for all simulations. Note that achieving these values for the Lundquist and magnetic-Reynolds numbers associated to the Laplacian resistivity has been only possible by simultaneously employing an eighth-order hyper-dissipation operator (whose coefficient has been carefully chosen following a detailed convergence study).

\subsection{Timescales of the problem}\label{subsec:timescales}

There are three important timescales that govern the dynamics of the cascade. The first is the interaction time defined by $\tau_{\rm int}^{-1}=(\tau_{\rm lin}^{+})^{-1}+(\tau_{\rm lin}^{-})^{-1}=(2\pi)^{-1}(k_{\|}^{+}v_{\rm A}+k_{\|}^{-}v_{\rm A})$, the time between two consecutive collisions of AW packets. 
In our setup, $\tau_{\rm int}=\tau_{\rm A}/3$, where $\tau_{\rm A}\doteq L_0/v_{\rm A}$ is the Alfv\'en crossing time. The second timescale is $\tau_{*}$, the time at which the turbulence reaches its ``peak activity'', estimated as a multiple $N_*$ of the nonlinear timescale $\tau_{\rm nl}\approx\tau_{\rm A}/\chi_0$. 
Note that smaller values of $\chi_0$ correspond to larger $\tau_{\rm nl}$. 
Usually, a few nonlinear times are required to reach a peak in the root-mean-square (rms) current density, $J_{\rm rms}$~\citep[e.g.,][]{ServidioNPG2011}; we find $N_*\approx 3$ in our simulations. 
Fully developed turbulence should thus be reached after $N^\ast_{\rm int} = \tau_*/\tau_{\rm int} \sim 9/\chi_0$ collisions (Figure~\ref{fig:Jrms_compare_hst}, left-panel inset, dotted line). 
This number is noticeably smaller than implied by standard weak-turbulence estimates\footnote{Even at $\chi_0\ll1$, our setup may not necessarily satisfy some working assumptions that are typical of standard weak-turbulence (WT) theory. WT theory assumes (weak) interactions between a sea of different, randomly phased waves, whereas in our simulations the (weak) interactions occur always between the same two (randomly phased) waves. Although the simulated turbulence is weak at large scales when $\chi_0<1$, this somewhat artificial setup may invalidate the ``random-walk argument'' leading to the $N^*\propto \chi^{-2}$ scaling in standard WT theory.}, for which $\tau_*$ would be a few cascade times; e.g., by analogy, if $\tau_*\sim 3\tau_{\rm casc}\approx 3\tau_{\rm nl}/\chi_0$, then $N^* \sim 9/\chi^2_0$ (inset, dashed line). The difference between the data and the weak-turbulence estimate motivates the introduction of a third timescale, the inverse growth rate of the tearing instability, $(\gamma^{\rm t})^{-1}$. 
If $\gamma_{k_*}^{\rm t}\tau_{{\rm nl},k_*}\gtrsim1$ at some scale $k_*$, the tearing instability is able to feed off of the associated current sheet in the cascade before the host eddy decorrelates through nonlinear interactions. 
Such a transition scale in the strong $\chi_0\sim1$ regime of MHD turbulence has been shown to scale as $k_*L_0\propto S_0^{4/7}$ by a number of authors~\citep[e.g.,][]{LoureiroBoldyrevPRL2017,MalletMNRAS2017,ComissoAPJ2018}. 
Because $\tau_{\rm nl}$ is larger for smaller $\chi_0$, this tearing condition should be easier to satisfy at larger scales (smaller $k_*$) for weak nonlinearities than in strong turbulence. 
The idea that a tearing-mediated range could emerge within a weakly nonlinear cascade also  relies implicitly on the fact that, analogously to what was postulated by \citet{Boldyrev2006} for strong turbulence, some sort of dynamic alignment of turbulent fluctuations occurs in the weak regime as well, so that the fluctuations become 3D anisotropic. 
In \S\ref{subsubsec:alignment_sim} we show that this is indeed the case, and that the observed scalings \citep[which differ significantly from those predicted for strong turbulence by][]{Boldyrev2006} can be explained by a phenomenological theory for dynamically aligned weak turbulence (\S\ref{subsec:theory_alignment}).
This argument is one motivation for our focus on $\chi_0 < 1$, since it implies that less numerical resolution is required to realize tearing-mediated turbulence at small $\chi_0$ than within a dynamically aligned, critically balanced state having $\chi_0 \sim 1$ (\S\ref{subsec:theory_reconnection_scale}).
We will additionally argue that CB is induced by reconnection in the tearing-mediated range, and that this may explain both the observed fluctuations' scaling in this range and the reduced number of AW-packets interactions, $N_{\rm int}^*\propto\chi_0^{-1}$ instead of $N_{\rm int}^*\propto\chi_0^{-2}$, needed to achieve the peak activity at low $\chi_0$ (\S\ref{subsec:theory_reconnection_CB}).

\begin{figure*}[!ht]
\centering
\includegraphics[width=0.48\textwidth]{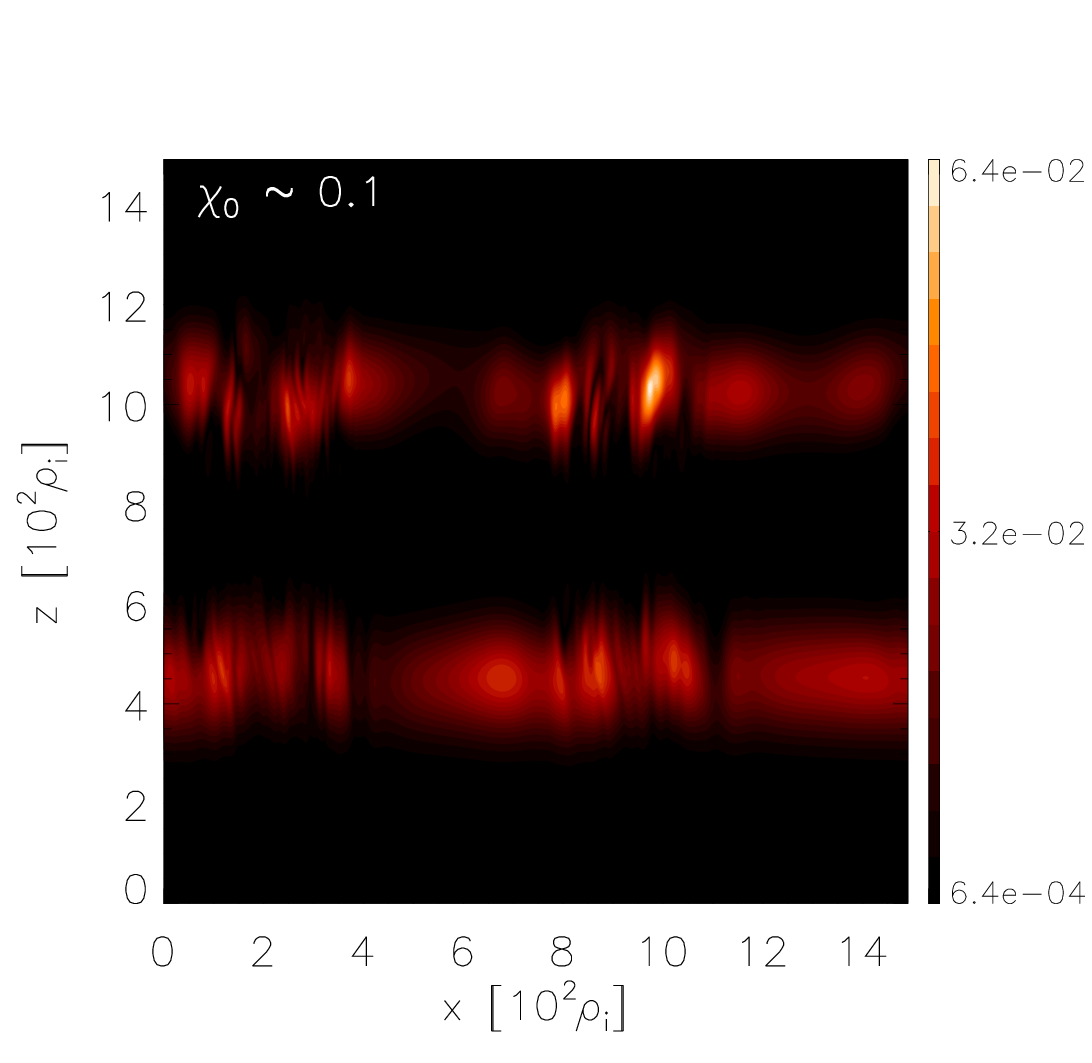}%
\includegraphics[width=0.48\textwidth]{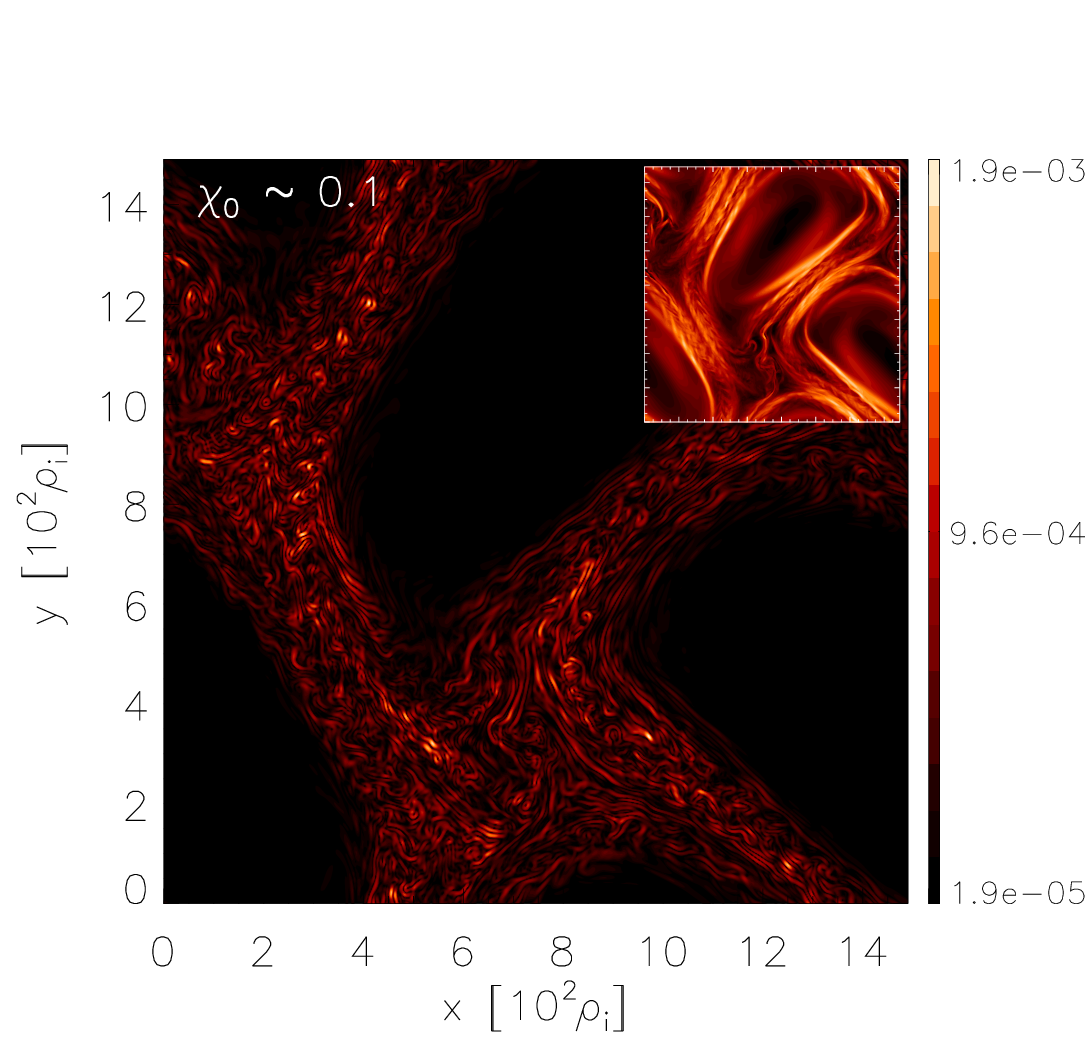}\\
\vspace{-1.33cm}%
\includegraphics[width=0.48\textwidth]{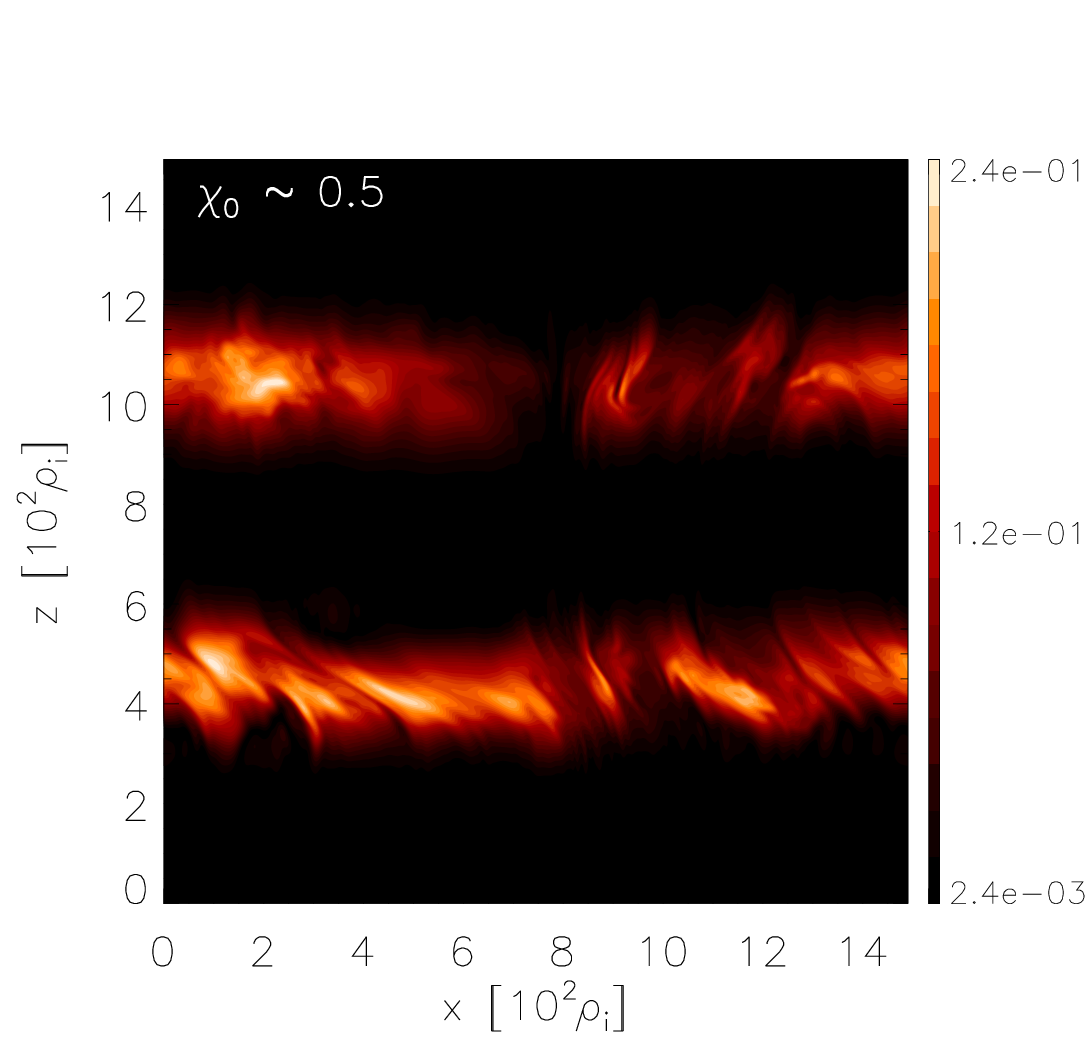}%
\includegraphics[width=0.48\textwidth]{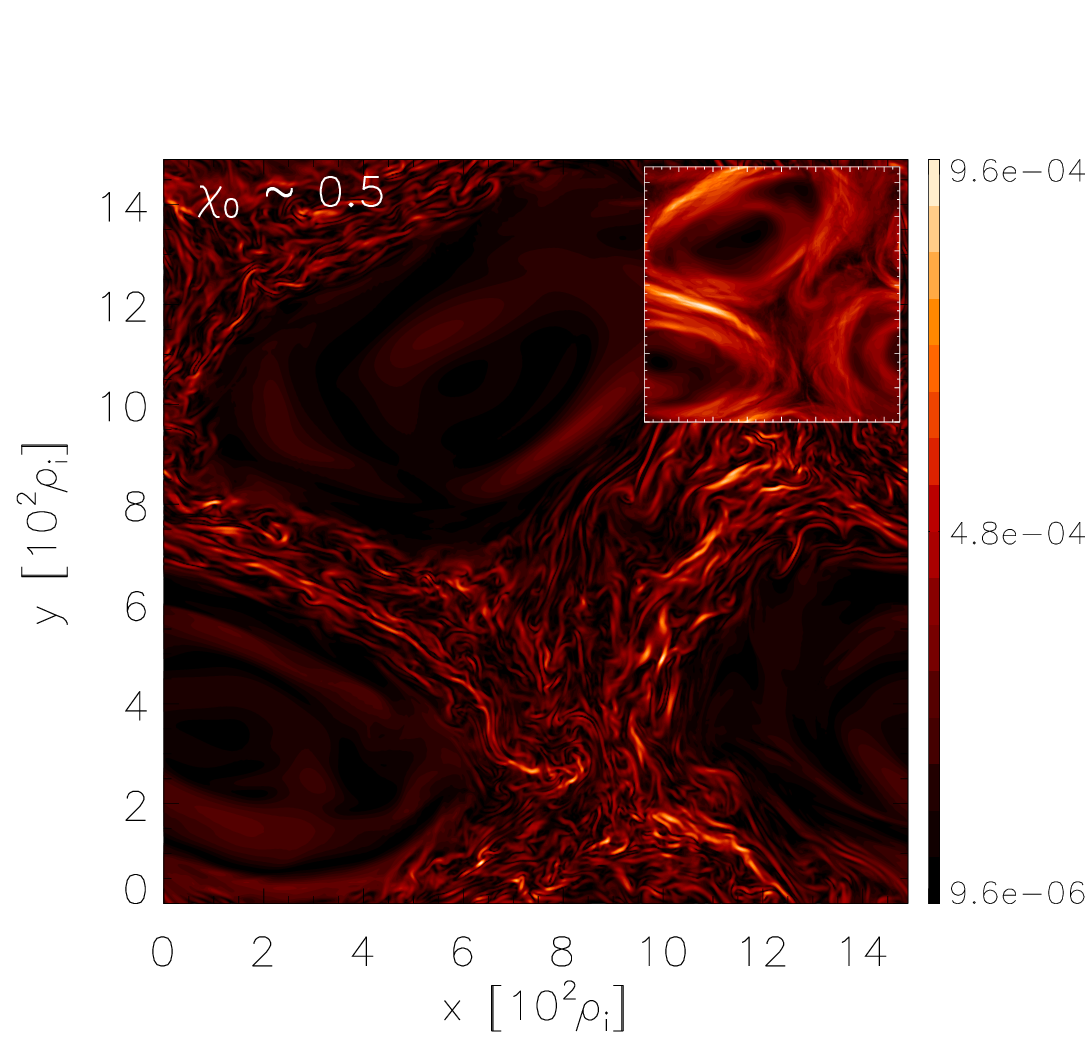}\\
\vspace{-1.33cm}%
\includegraphics[width=0.48\textwidth]{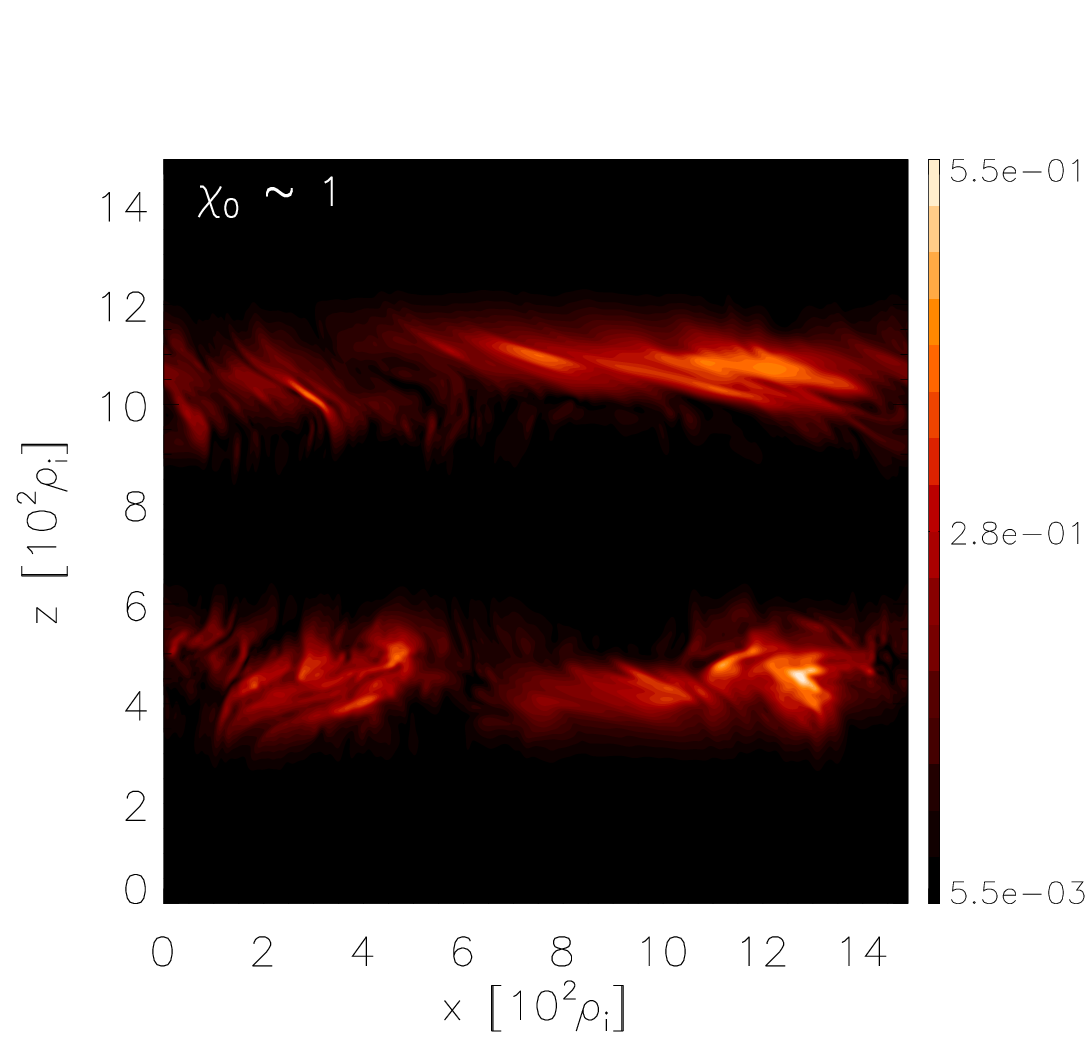}%
\includegraphics[width=0.48\textwidth]{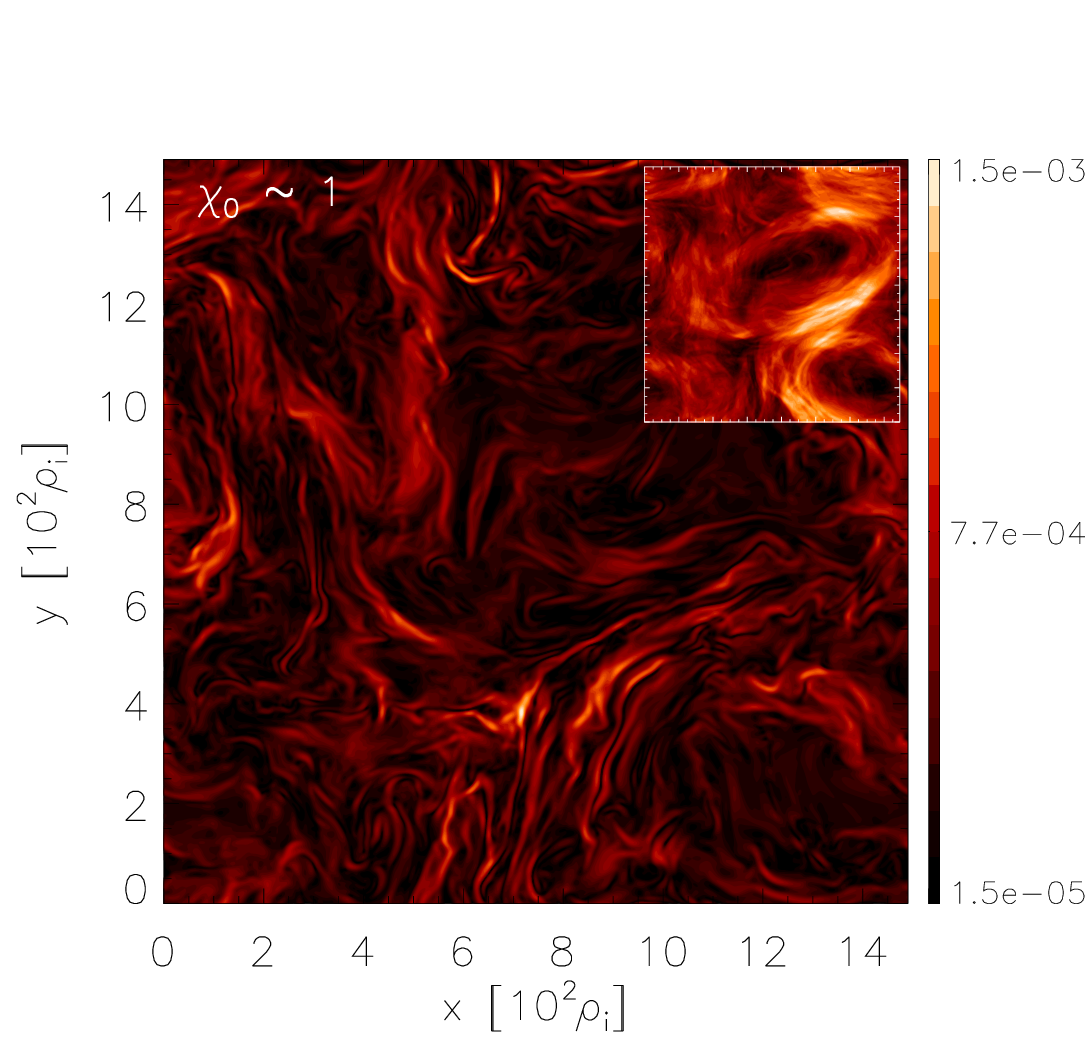}\\
\caption{Iso-contours of $\delta B_\perp/B_{\rm rms}$ in the $x$-$z$ plane at $y=L_0/2$ (left column) and in the $x$-$y$ plane at $z=L_0/2$ (right column), at $t/\tau_{*}\simeq1.35$ (in the developed turbulent regime) for initial non-linearity parameter $\chi_0\sim0.1$, ${\sim}0.5$ and ${\sim}1$ (from top to bottom). Recall that $\bb{B}_0$ is along $z$. Insets: iso-contours of $\delta B_\perp/B_{\rm rms}$ averaged over $z$. (see \href{https://lagrange.oca.eu/fr/silvio-cerri/3794-animations}{webpage of S.~S.~Cerri} for animations.)}
\label{fig:Bperp_contour_xy}  
\end{figure*}

\section{Numerical results}\label{sec:results}

Simulations are performed for a few $\tau_{*}$ (Figure~\ref{fig:Jrms_compare_hst}), corresponding to a large number of AW-packet collisions (e.g., $N_{\rm int}^{\rm(tot)}\approx200$ at $\chi_0\sim0.1$). 
If not stated otherwise, fluctuations' properties are determined by averaging over a time interval $\Delta t\approx0.8\tau_*$ around peak-activity.

\begin{figure*}[!ht]
\center%
\includegraphics[width=0.5\textwidth]{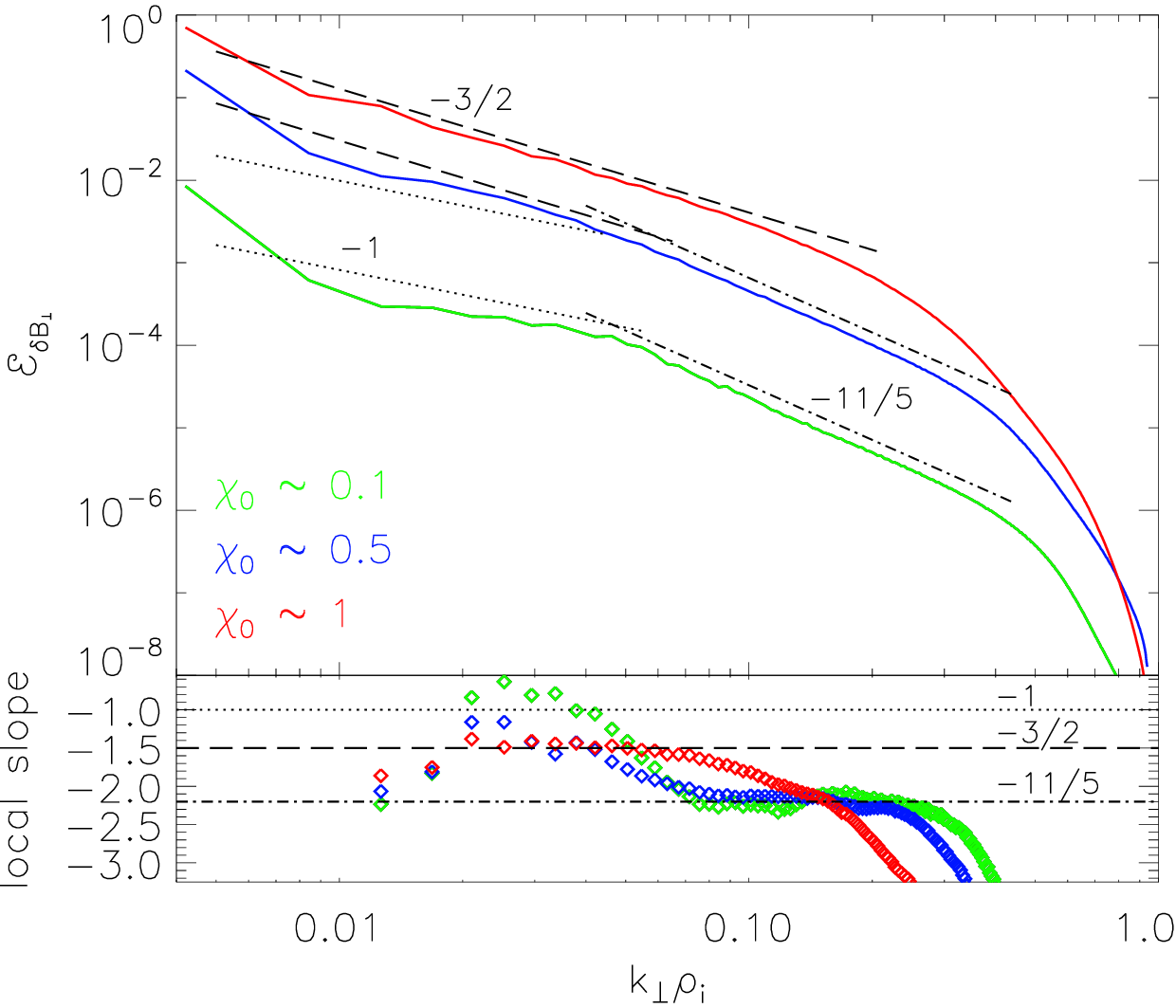}%
\includegraphics[width=0.5\textwidth]{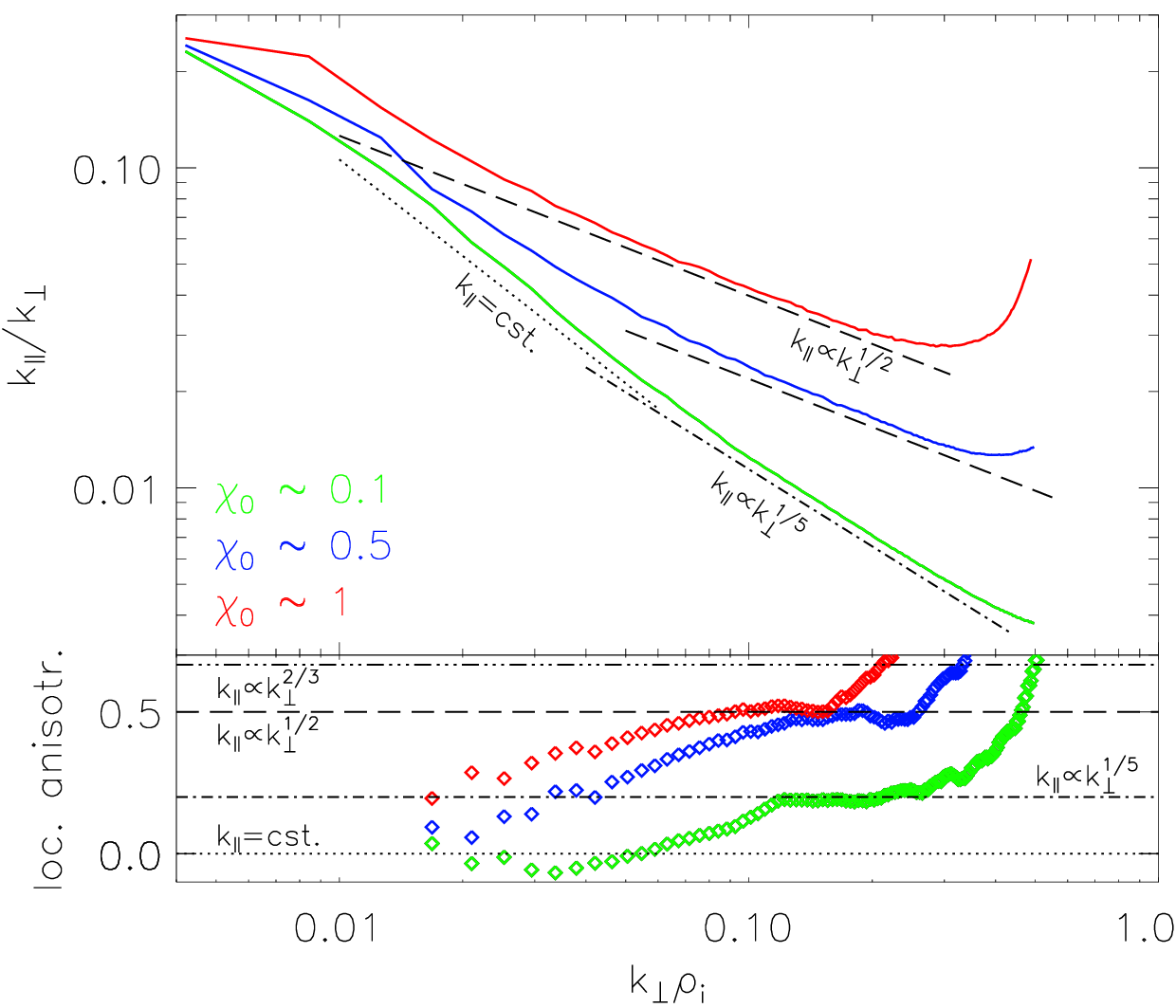}%
\caption{Left: $\delta B_\perp$ energy spectrum and its local  slope versus $k_\perp\rho_{\rm i}$. Spectra are time-averaged over $0.9\lesssim t/\tau_*\lesssim 1.7$. Right: spectral anisotropy $k_\|/k_\perp$ averaged over the same time interval [$k_\|(k_\perp)$ is obtained using the method presented in \citet{ChoAPJL2002}]. Relevant power laws are provided for reference.}
\label{fig:spectra_compare}  
\end{figure*}

\subsection{Fluctuations' properties at peak activity}\label{subsec:analysis_peak}

As AW packets shear one another in the plane perpendicular to $\bb{B}_0$, they generate strong CSs (evidenced by the short-time oscillations in $J_{\rm rms}$; Figure~\ref{fig:Jrms_compare_hst}).  
Each interaction increases the magnetic shear in the CSs, thus increasing $J_{\rm rms}$ until ``peak activity'' is eventually achieved.

\subsubsection{Current-sheet disruption and AW-packets' structure}\label{subsubsec:peak_structures}

Figure~\ref{fig:Bperp_contour_xy} shows perpendicular magnetic-field fluctuations, $\delta B_\perp/B_{\rm rms}$, both in the $x$-$z$ plane (left column) and in the $x$-$y$ plane (right column) at $t/\tau_{*}\simeq1.35$, after turbulence has developed.
At this time, AW packets are still clearly distinguishable in the $x$-$z$ plane (left column), with more fine-scale structure visible within the packets with increasing $\chi_0$ (top to bottom). 
These are related to CS structures formed through AW-packet collisions~\citep[e.g.,][]{PezziJPP2017,VernieroJPP2018a}, which are then affected by tearing instability occurring within them.
At $\chi_0\sim0.1$, they are well localized in $x$ and have essentially no structure along $z$ (Figure~\ref{fig:Bperp_contour_xy}, top left panel). 
The occurrence of finer structures along $z$ (corresponding to the generation of small-$k_\|$ scales; see \S\ref{subsubsec:peak_spectra}) increases with increasing $\chi_0$ (Figure~\ref{fig:Bperp_contour_xy}, middle left and bottom left panels, respectively).
While the bulk of the AW packets are still distinguishable in all regimes, the structure of $\delta B_\perp$ in the plane perpendicular to $\bb{B}_0$ exhibits clear differences. 
The ``relics'' of disrupted CSs are especially recognizable at $\chi_0\sim0.1$, where $\delta B_\perp$ fluctuations indeed resemble small-scale, plasmoid-like structures in 2D (i.e., quasi-circular magnetic structures referred to as ``magnetic islands'' in 2D, which in 3D actually manifest as flux ropes; Figure~\ref{fig:Bperp_contour_xy}, top right panel). 
At $\chi_0\sim0.5$, such structures are also visible, although $\delta B_\perp$ fluctuations are now less organized into plasmoid-like structures within the disrupted CSs (this difference reflects on the low-$k_\perp$ part of the $\delta B_\perp$ energy spectrum; see \S\ref{subsubsec:peak_spectra}). 
$\delta B_\perp$ fluctuations are clearly different at $\chi_0\sim1$, where no large-scale CS structures are distinguishable in the perpendicular plane (Figure~\ref{fig:Bperp_contour_xy}, bottom right panel): this is qualitatively similar to 3D turbulence arising from broad-band injection~\citep[see, e.g., Figure~1 of][]{CerriFSPAS2019}.

\subsubsection{Fluctuations' spectrum and anisotropy}\label{subsubsec:peak_spectra}

As a result of AW interaction and CS disruption, a cascade of $\delta B_\perp$ fluctuations develops (Figure~\ref{fig:spectra_compare}). 
At $\chi_0\sim0.1$ and ${\sim}0.5$ (green and blue curves, respectively), the $\delta B_\perp$ energy spectra exhibit a break at $k_\perp\rho_{\rm i}\approx0.05$ (Figure~\ref{fig:spectra_compare}, top-left panel), which we identify as the transition scale $k_*$. 
Both simulations indeed show a ``small-scale'' MHD spectrum below $k_*$ proportional to $k_\perp^{-\alpha}$ with spectral index $2.1\lesssim\alpha\lesssim2.3$ (Figure~\ref{fig:spectra_compare}, bottom-left panel), consistent with predictions for tearing-mediated turbulence~\citep[{\em viz.}, between $k_\perp^{-11/5}$ and $k_\perp^{-19/9}$; see, e.g.][]{MalletMNRAS2017,BoldyrevLoureiroAPJ2017,ComissoAPJ2018,TeneraniVelli2020}. 
Such a spectral break is instead not present in the $\chi_0\sim1$ case, consistent with the expectation that strong turbulence would require a larger $S_0$ to resolve $k_*$ (see \S\ref{subsec:theory_reconnection_scale}). At $k_\perp\rho_{\rm i}\lesssim0.05$, however, the two regimes develop a different power law (although of limited extent), close to $-3/2$ at $\chi_0\sim0.5$ and to $-1$ at $\chi_0\sim0.1$.
Although it would be appealing to interpret the $-3/2$ spectrum within the context of a dynamically aligned, strong MHD turbulent cascade~\citep[][]{Boldyrev2006,ChandranAPJ2015,MalletSchekochihinMNRAS2017}, we found $\chi_k<1$ at $k_\perp<k_*$ (not shown). Analogously, the $-1$ spectrum may be due to a not-yet-developed large-scale turbulent state, or perhaps to non-local transfer between the AW packets and the disruption scale through CS structures~\citep[cf. Figure 4 in][]{FranciAPJL2017}.
Nevertheless, fluctuations at both $\chi_0\sim0.1$ and ${\sim}0.5$ show a spectral anisotropy $k_\parallel/k_\perp$ consistent with the weak-turbulence regime at $k_\perp<k_*$ (i.e., $k_\parallel\approx{\rm const}$; Figure~\ref{fig:spectra_compare}, right panel).
A possible alternative explanation for the above spectra in terms of dynamic alignment in weak turbulence is provided in \S\ref{sec:theory}.
On the other hand, the formation of a $k_\perp^{-3/2}$ spectrum at $\chi_0\sim1$ (Figure~\ref{fig:spectra_compare}, left panels, red curve) is consistent with dynamic alignment in strong MHD turbulence. This seems to be confirmed by the measured spectral anisotropy $k_\parallel\propto k_\perp^{1/2}$ (Figure~\ref{fig:spectra_compare}, right panel).

\subsubsection{Fluctuations' alignment angle}\label{subsubsec:alignment_sim}

As anticipated in \S\ref{subsec:timescales}, the possibility to activate a tearing-mediated cascade relies not only on the fact that turbulent eddies are sheared in the field-perpendicular direction to set up a tearing-unstable configuration, but also on the requirement that these eddies live long enough to allow tearing instability to grow and disrupt them. The former is a consequence of the dynamic alignment of turbulent fluctuations in that plane, which ultimately gives the fluctuations a 3D spectral anisotropy. However, once reconnection sets in, the effect of the eddies' disruption by the tearing instability is to interrupt the achieved cascade-induced alignment by producing plasmoid-like structures (i.e., replacing the elongated sheet-like structure of the eddy in the field-perpendicular plane with quasi-circular magnetic islands---flux ropes, in 3D); this process instead {\em increases} the alignment angle~\citep[i.e., produces ``misalignment''; see, e.g.][]{MalletMNRAS2017,BoldyrevLoureiroAPJ2017,ComissoAPJ2018}. This is interpreted by \cite{MalletMNRAS2017} in terms of a discrete and recursive view of the cascade: once the cascade enters the tearing-mediated range $\lambda\lesssim\lambda_*$, there will be a ``reset'' of the fluctuations' alignment angle and amplitude due to the eddy disruption---increasing the former and decreasing the latter---followed by a range in which these fluctuations  cascade further towards smaller scales while re-aligning until the condition for tearing-induced disruption is achieved, again ``resetting'' the alignment and amplitude, and so on until dissipation sets in (see discussion in their \S 6). On the other hand, \citet{BoldyrevLoureiroAPJ2017} and \citet{ComissoAPJ2018} assume that, below $\lambda_*$, the fluctuations will keep mis-aligning with decreasing scale, with the scaling $\sin\theta_\lambda\propto\lambda^{-4/5}$. 
This scaling is based solely on the physics of tearing instability, i.e., on the scalings of the nonlinear Coppi mode~\citep{CoppiFizPl1976}, and thus formally belongs to a pure tearing-mediated cascade (i.e., occurring homogeneously in space and time). 
The recursive-disruption view of \citet{MalletMNRAS2017} also produces a scale-dependent alignment angle, which is constrained within an envelope whose boundary scales as $\lambda^{-4/5}$~\citep[see \S 7.2.3 in][]{Schekochihin_arxiv2020}. This $4/5$ envelope can be interpreted as the strongest alignment sustainable in the tearing-mediated range, but it is not clear {\em a priori} what would emerge as a global feature in $k$ space (i.e., resulting from a spatial average). 
The main difference between these two views depends on the details behind the X-point collapse~\citep[for a detailed discussion, see \S 7.4.1 in][]{Schekochihin_arxiv2020}.
Although these two interpretations are not incompatible in term of the resulting fluctuations' spectrum, they could differ in terms of the effective scale-dependent alignment angle that can be {\em measured}.

We offer here a different point of view, somewhat {\em complementary} to the two summarized above. One can actually think about the ensemble of turbulent fluctuations as dynamically aligning (via the usual, non-tearing-mediated cascade) and mis-aligning (through tearing) in a patchy fashion in space and time, rather than step-wise in $k$ space. This will likely result in a complicated convoluted state, when globally averaged over the ensemble (i.e., not necessarily providing a clean, global $k_\perp^{4/5}$ scaling). 
Here, we illustrate this patchy-in-time behavior by distinguishing between those periods when the AW packets are shearing one another during their interaction (``overlap'') and those periods during which the AW packets are instead far apart (``free cascade''). This will demonstrate that both states, a dynamically aligned cascade and a tearing-mediated mis-aligning cascade, are recursively realized in time. On the other hand, by performing the same time average as done for the fluctuations' spectra (which are indeed not affected by distinguishing between the above stages), we have found that $\sin\theta_\lambda$ exhibits ambiguous scalings (not shown). For our setup, {\em viz.}~AW-packet collisions, taking into account the patchiness in space appears to be less important (especially in the $\chi_0\sim0.1$ regime, where the largest-scale fluctuations affect much less the tearing-mediated regions). 
However, we expect that in simulations with broad-band injection, this spatial patchiness should be carefully taken into account in order to capture the correct scalings of mis-alignment in the tearing-mediated range.

\begin{figure*}[!ht]
\center%
\includegraphics[width=0.33\textwidth]{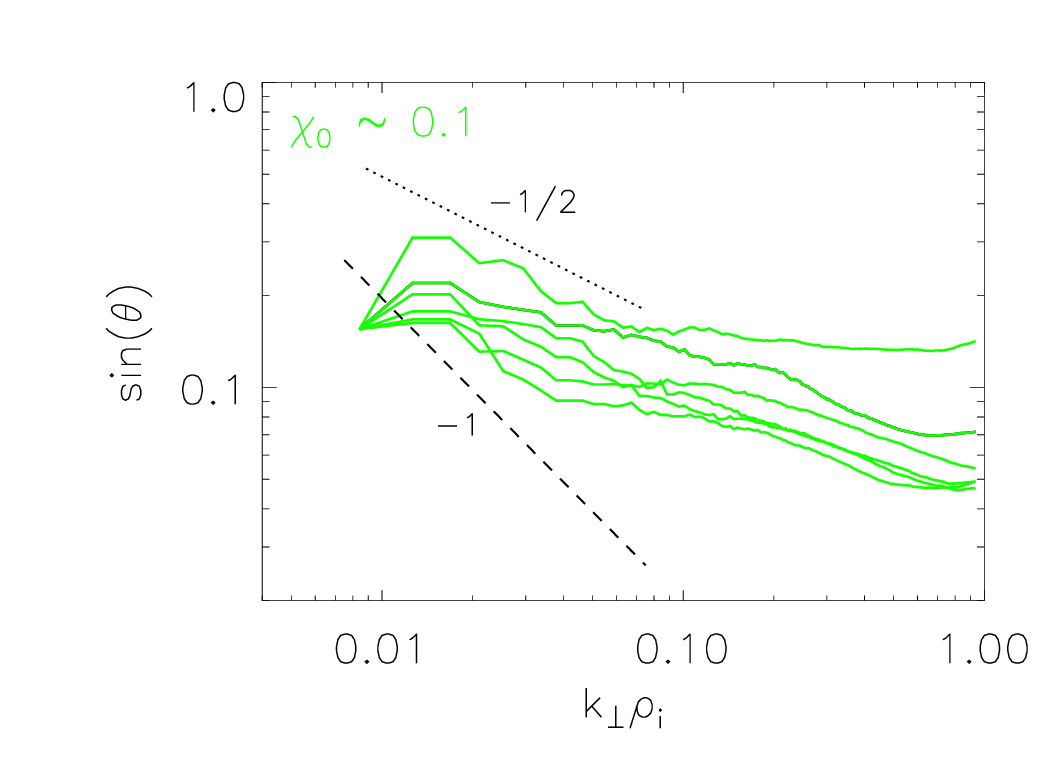}%
\includegraphics[width=0.33\textwidth]{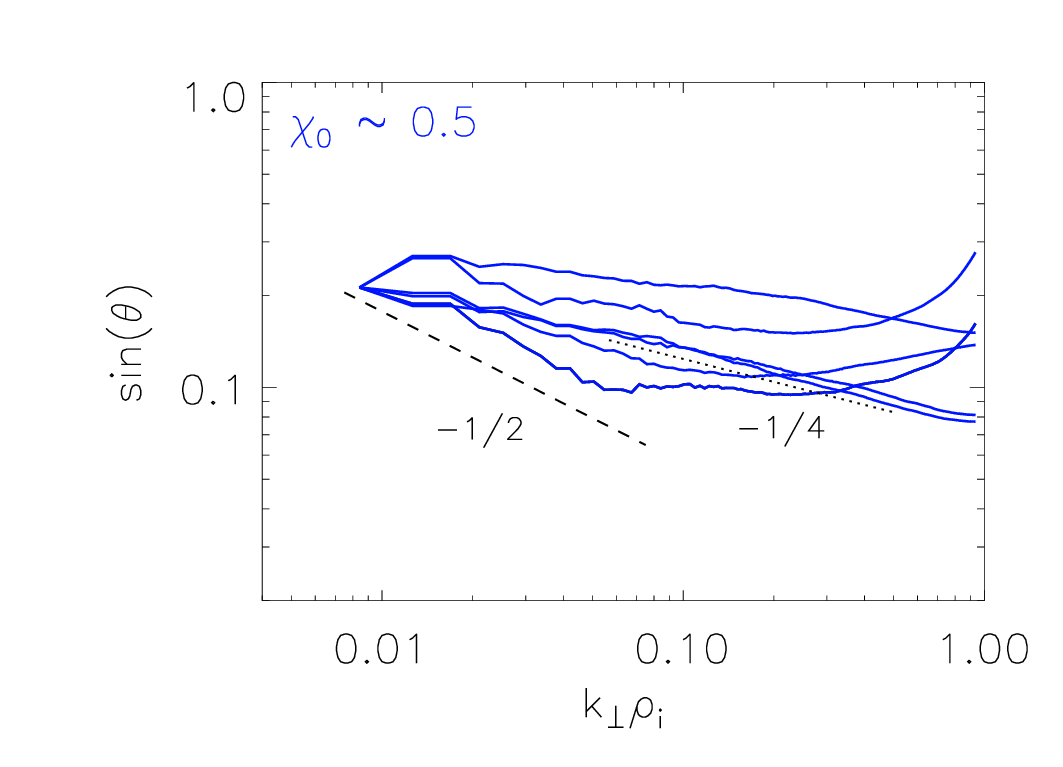}%
\includegraphics[width=0.33\textwidth]{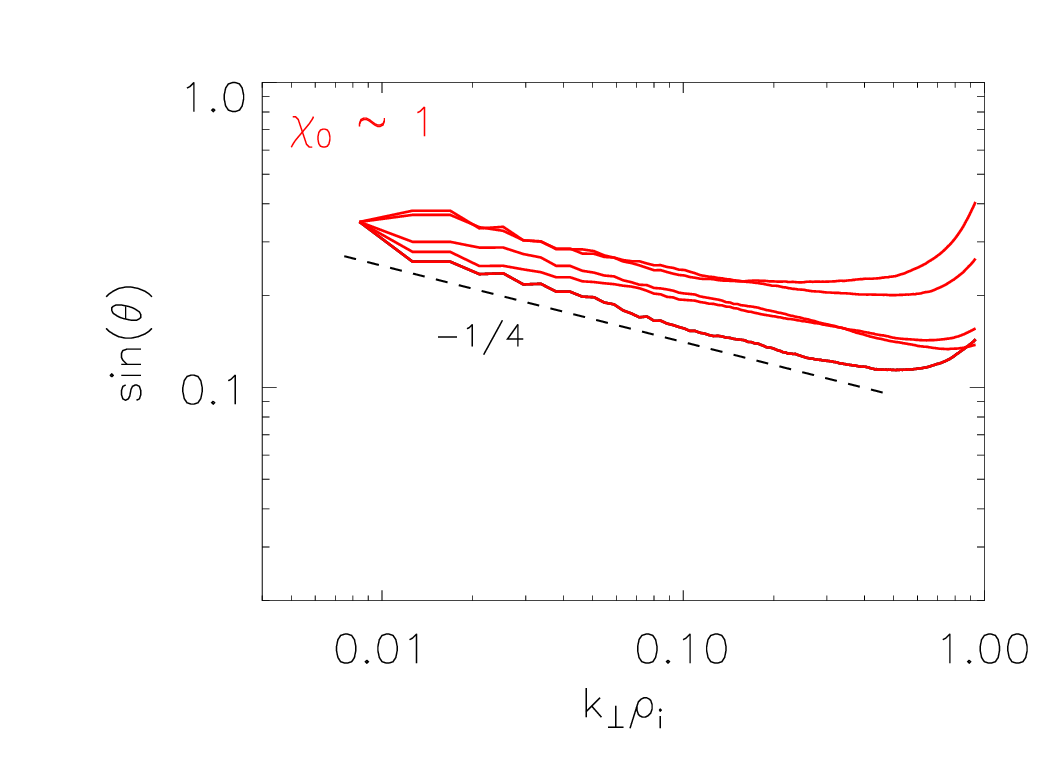}\\
\includegraphics[width=0.33\textwidth]{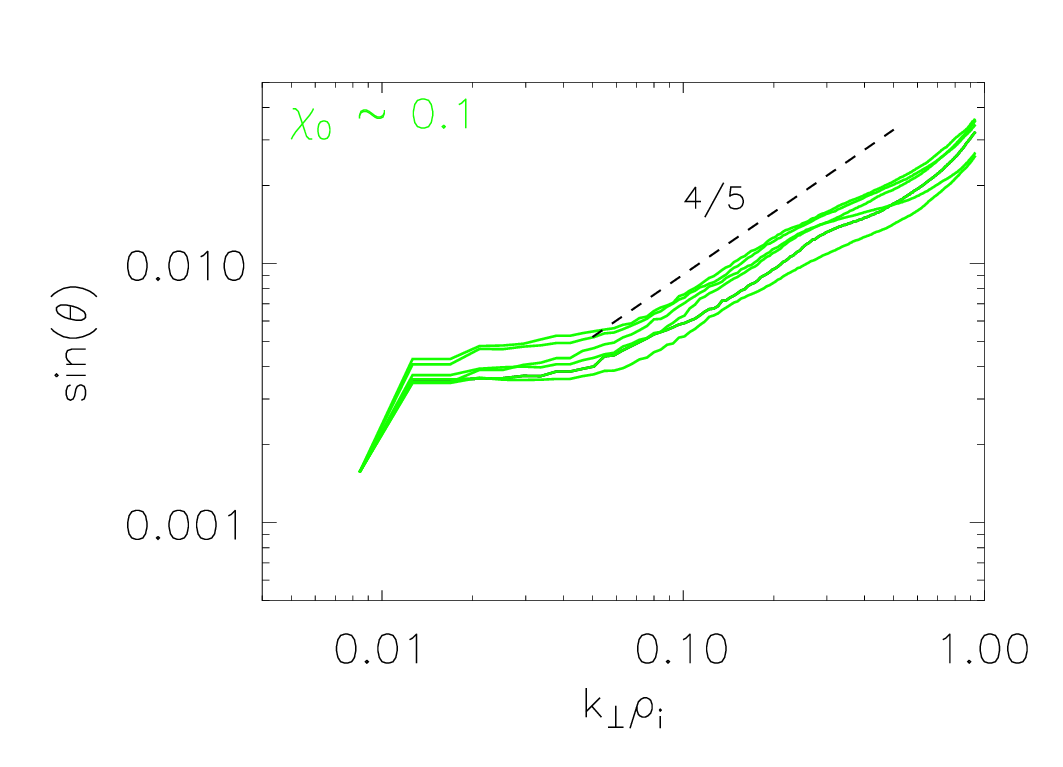}%
\includegraphics[width=0.33\textwidth]{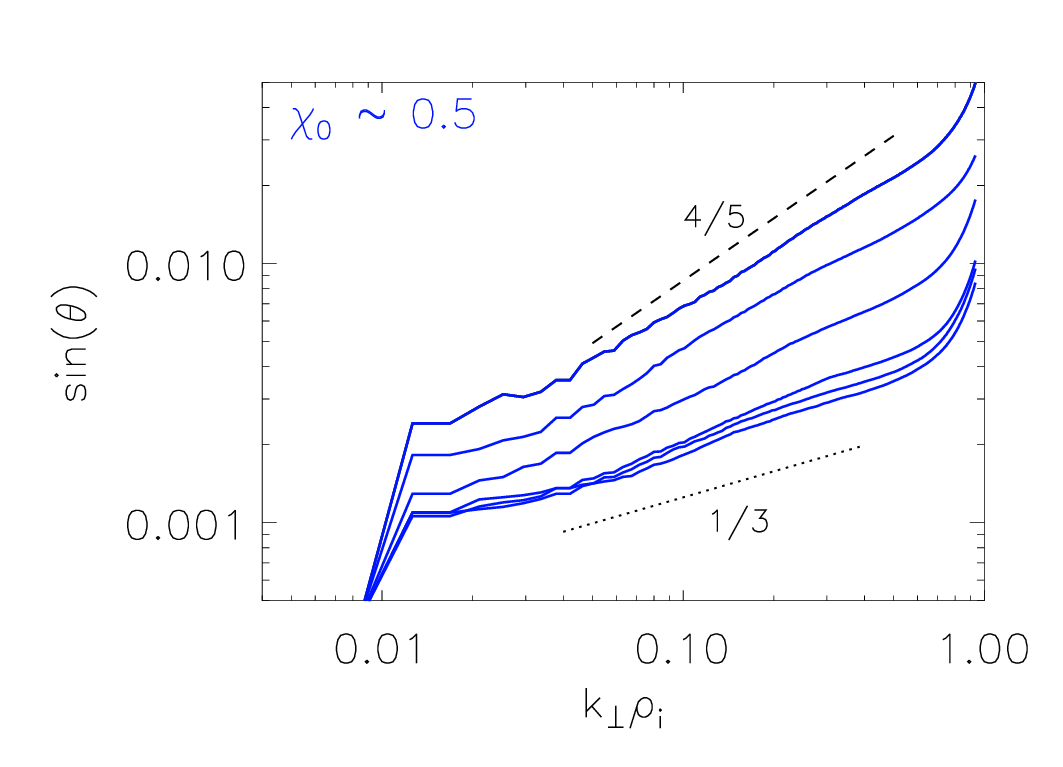}%
\includegraphics[width=0.33\textwidth]{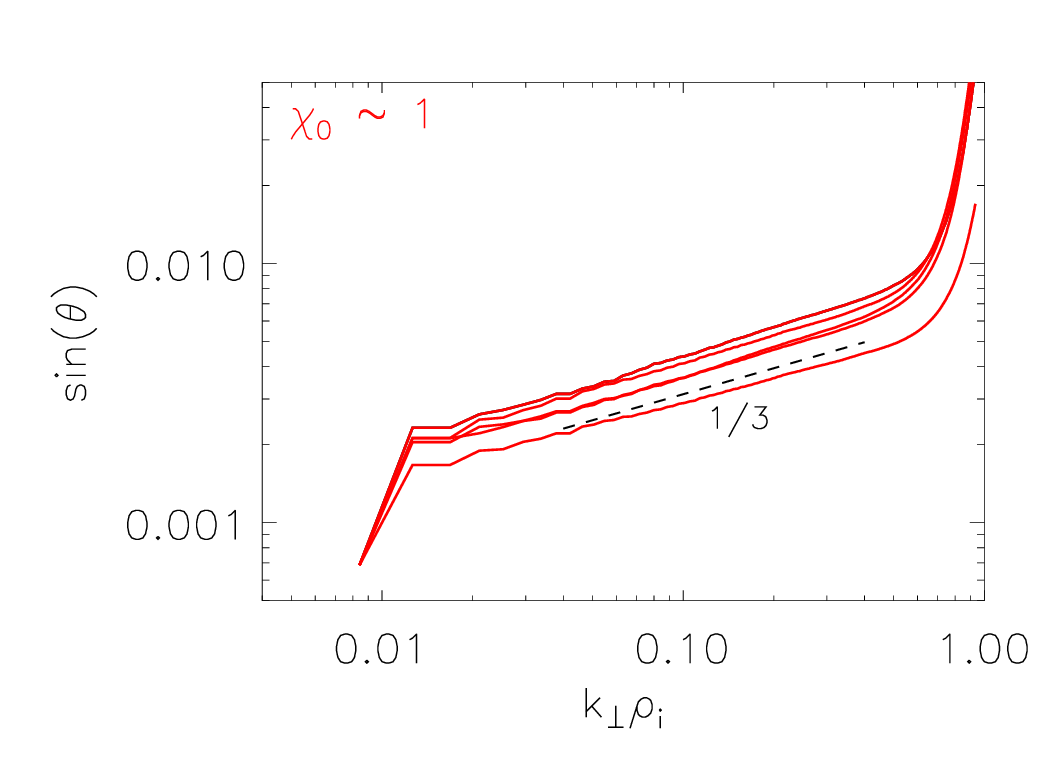}%
\caption{Variation of the alignment angle with scale for different initial values of the nonlinearity parameter $\chi_0$ and at different times within developed turbulence ($0.9\lesssim t/\tau_*\lesssim 1.7$): $\sin(\theta)$ versus $k_\perp\rho_{\rm i}$ at times when AW packets spatially overlap (top row) and when they are far apart (bottom row). Scale-dependent alignment is computed using Equation~\eqref{eq:alignment_def}. Relevant power laws are provided for reference.}
\label{fig:alignment}  
\end{figure*}

Analogously to the calculation of wavenumber anisotropy~\citep{ChoAPJL2002}, we estimate the alignment angle $\theta$ between the velocity- and magnetic-field fluctuations at $k_\perp$ using\footnote{In order to  estimate  $\sin\theta_{k_\perp}$ correctly, it is important to employ the averaging procedure $\langle|\delta\bb{u}_{\perp,\lambda}\btimes\delta\bb{b}_{\perp,\lambda}|\rangle/\langle|\delta\bb{u}_{\perp,\lambda}||\delta\bb{b}_{\perp,\lambda}|\rangle$ instead of a normalized version $\langle|\delta\bb{u}_{\perp,\lambda}\btimes\delta\bb{b}_{\perp,\lambda}|/(|\delta\bb{u}_{\perp,\lambda}||\delta\bb{b}_{\perp,\lambda}|)\rangle$. This is needed to select the ``dynamically relevant'' fluctuations; i.e., the averaging procedure should reflect the fact that, at a given scale $\lambda$, the fluctuations that contribute the most to the turbulent dynamics are those whose amplitudes are close to the rms value at that scale; see discussion in \citet{MasonPRL2006}.}
\begin{equation}\label{eq:alignment_def}
    \sin \theta_{k_\perp}\,=\,
    \frac{\langle\sum_{k\leq {k}_\perp<k+1}|\delta\bb{u}_{\perp,\lambda}\btimes\delta\bb{b}_{\perp,\lambda}|\rangle}{\langle\sum_{k\leq {k}_\perp<k+1}|\delta\bb{u}_{\perp,\lambda}||\delta\bb{b}_{\perp,\lambda}|\rangle}\,,
\end{equation}
where the perpendicular direction is defined with respect to a scale-dependent mean field, $\langle\bb{B}\rangle_\lambda$, obtained by eliminating modes with $k_\perp>k/2\sim1/(2\lambda)$ from $\bb{B}$.
Results for different simulations are shown in Figure~\ref{fig:alignment}, where we distinguish between the two main phases discussed above. 
During the interaction of the AW packets (``overlap''; top row), fluctuations get highly sheared and $\sin\theta_{k_\perp}$ clearly shows their tendency to align at decreasing scales. In the $\chi_0\sim1$ case, fluctuations align such that $\sin\theta_{k_\perp}\sim k_\perp^{-1/4}$, which matches the prediction by \citet{Boldyrev2006}. 
Note that, although this simulation seems to support the idea that dynamic alignment in strong turbulence is an effect that proceeds all the way down to dissipation scales~\citep[as observed by][]{PerezPRX2012}, the limited resolution in our simulations cannot exclude the possibility that alignment is a finite-range effect that is tied to the dynamics at the outer scale and thus might stop before the dissipation scales are reached in the cascade~\citep[as claimed by][]{BeresnyakMNRAS2012}.
The cases with smaller values of $\chi_0$, on the other hand, exhibit stronger alignment: roughly as $k_\perp^{-1/2}$ for $\chi_0\sim0.5$ (perhaps reducing to $k_\perp^{-1/4}$ at the smallest scales, also showing small-scale flattening in some cases), and something in between $k_\perp^{-1}$ and $k_\perp^{-1/2}$ for $\chi_0\sim0.1$ (also exhibiting small-scale flattening in some cases). This behavior may be explained by the theory presented in \S\ref{subsec:theory_alignment}.
When AW packets are instead far apart (``free cascade''; bottom row), the fluctuations' dynamics is dominated by the tearing-mediated cascade and $\sin\theta_{k_\perp}$ exhibits a tendency to misalign. However, while the $\chi_0\sim0.1$ regime misaligns fluctuations roughly as $k_\perp^{4/5}$ for $k_\perp>k_*$ in all cases \citep[as predicted by][]{BoldyrevLoureiroAPJ2017}, the intermediate $\chi_0\sim0.5$ case also shows times with a slightly weaker misalignment at $k_\perp>k_*$ in addition to the $4/5$ scaling (somewhat between $k_\perp^{3/5}$ and $k_\perp^{1/2}$). This weaker dependence of the alignment angle at $\chi_0\sim0.5$ can be interpreted as the effect of some non-negligible amount of the spatial patchiness discussed above, which is present in this regime despite our simple set up of AW-packet collisions (see Figure~\ref{fig:Bperp_contour_xy}, middle row). During this ``relaxation'' stage, we also observe a weak misalignment for $\chi_0\sim1$, following approximately $k_\perp^{1/3}$ (without any obvious spectral breaks). We do not have any obvious explanation for this behavior at the moment, and further investigation would be required to address this point.

\section{Dynamic alignment and reconnection in weak Alfv\'enic turbulence}\label{sec:theory}

Dynamic alignment of turbulent fluctuations is a necessary condition for the cascade to realize a tearing-mediated regime. To explain the evidence for this regime occurring in our simulations, we take a step back and postulate how dynamic alignment would affect the standard weak-turbulence phenomenology.

In this Section, we provide a phenomenological description of weak turbulence in which dynamic alignment is occurring, and discuss its implications for possible transitions to CB and/or to tearing-mediated turbulence. 
For this purpose, we first establish our notation.  
Let us call $\lambda$ the perpendicular length of fluctuations in the direction perpendicular to both the mean magnetic field at such scale, $\langle\bb{B}\rangle_\lambda$, and the perpendicular (to $\langle\bb{B}\rangle_\lambda$) magnetic-field fluctuations $\delta\bb{b}_{\perp,\lambda}$ (in Alfv\'enic units). Then, $\ell_\lambda$ and $\xi_\lambda$ are the lengths of such fluctuations along $\langle \bb{B}\rangle_\lambda$ and along $\delta\bb{b}_{\perp,\lambda}$, respectively.\footnote{This distinction between the two scale-dependent transverse directions $\lambda$ and $\xi$ is neglected in Equation~\eqref{eq:alignment_def}, consistent with the assumption that angular spectral averaging makes the difference between the variation scales transverse to the local field and the ambient field subdominant.}
 Quantities evaluated at the injection scale are adorned with a ``0'' subscript. 
Following Fig.~3 of \citet{Boldyrev2006}, we define $\theta_\lambda$ as the angle between the flow- and magnetic-field fluctuations perpendicular to $\langle\bb{B}\rangle_\lambda$ at scale $\lambda$, $\delta\bb{u}_{\perp,\lambda}$ and $\delta\bb{b}_{\perp,\lambda}$, respectively. At the same time, if $\langle\bb{B}\rangle_\lambda$ differs from $\bb{B}_0$ by an angle $\widetilde{\theta}_\lambda$, the angle between $\delta\bb{u}_\lambda$ and $\langle\bb{B}\rangle_\lambda$ is $\pi/2-\widetilde{\theta}_\lambda$. These two angles scale as $\theta_\lambda\sim\lambda/\xi_\lambda$ and $\widetilde{\theta}_\lambda\sim\xi_\lambda/\ell_\lambda$, the total alignment angle between $\delta\bb{u}_\lambda$ and $\delta\bb{b}_\lambda$ being $\phi_\lambda^{ub}\simeq(\theta_\lambda^2+\widetilde{\theta}_\lambda^2)^{1/2}$~\citep[see \S2 of][]{Boldyrev2006}.
In general, dynamic alignment weakens non-linear interactions, $\mathcal{N}_\lambda\sim(\delta\bb{z}^{+}\bcdot\grad)\delta\bb{z}^{-}\sim(\delta\bb{z}^{-}\bcdot\grad)\delta\bb{z}^{+}\sim\phi_\lambda\delta z_\lambda^2/\lambda$, and simultaneously increases the cascade time $\tau_\lambda\sim \tau_{{\rm nl},\lambda}^2/\tau_{\rm A}\sim\tau_{\rm A}\phi_\lambda^{-2}(\lambda/\ell_0)^2(\delta z/v_{\rm A})^{-2}$, where now $\phi_\lambda$ is the angle between $\delta\bb{z}_\lambda^+$ and $\delta\bb{z}_\lambda^-$, $\tau_{\rm A}=\ell_0/v_{\rm A}$ and $v_{\rm A}=v_{\rm A,0}=\mathrm{const}$.\footnote{In general, it is not obvious whether one should define the alignment angle with respect to the fluctuations $\delta\bb{u}_\lambda$ and $\delta\bb{b}_\lambda$ or to the Els\"asser fields $\delta\bb{z}_\lambda^+$ and $\delta\bb{z}_\lambda^-$. In fact, while both the original theory by \citet{Boldyrev2006} and a number of  {\em in-situ} spacecraft measurements and of simulations' analyses focus on the former, showing the tendency of $\delta\bb{u}_\lambda$ and $\delta\bb{b}_\lambda$ to align with decreasing scales~\citep[e.g.,][]{MasonPRL2006,MasonAPJ2011,MatthaeusPRL2008,PodestaJGR2009,HnatPRE2011,PerezPRX2012}, it should be the latter that directly enters the nonlinear term in the Els\"asser formulation of the MHD equations~\citep[i.e., it is the $\delta\bb{z}^\pm$'s that shear one another into alignment; see, e.g.,][]{BeresnyakLazarianAPJ2006,BeresnyakMNRAS2012,ChandranAPJ2015,MalletSchekochihinMNRAS2017}. Nevertheless, dynamic alignment of both $\delta\bb{u}_\lambda$ and $\delta\bb{b}_\lambda$, and of $\delta\bb{z}_\lambda^+$ and $\delta\bb{z}_\lambda^-$ are indeed simultaneously taking place~\citep[e.g.,][]{WicksPRL2013,MalletMNRAS2016}. The angles between the two set of fields are ultimately related by cross-helicity and residual energy, and both angles scale with $\lambda$ in the same way under certain circumstances~\citep[see, e.g.,][for a more detailed discussion on this matter]{Schekochihin_arxiv2020}.} 
[Note that, while we use the angle $\phi_\lambda$ (or, $\phi_\lambda^{ub}$) in the phenomenological scaling, it is actually $\sin\phi_\lambda$ (or, $\sin\phi_\lambda^{ub}$) that enters the nonlinear term, so that its effect on nonlinearities is symmetric with respect to the fact that $\delta\bb{z}_\lambda^+$ and $\delta\bb{z}_\lambda^-$ (or, $\delta\bb{u}_\lambda$ and $\delta\bb{b}_\lambda$) can either align or counter-align.]
In the following, we assume balanced turbulence at large scales\footnote{Assuming balance at large scales does not imply that a scale-dependent imbalance and residual energy is not present, and actually it can be seen from simple geometrical arguments that dynamic alignment indeed requires that both develop along the cascade.}, $|\delta\bb{z}^+|^2\approx|\delta\bb{z}^-|^2$, so that $\phi_\lambda^{ub}$ scales as $\phi_\lambda$, and we use the alignment angle $\theta_\lambda$ between $\delta\bb{u}_{\perp,\lambda}$ and $\delta\bb{b}_{\perp,\lambda}$ fluctuations as the relevant angle in the following phenomenological scalings. In fact, we will see that the scaling $\phi_\lambda^{ub}\sim\theta_\lambda$ holds in all cases of interest. Moreover, $\theta_\lambda$ is the angle most relevant for the cascade of $\delta\bb{b}$ and $\delta\bb{u}$ fluctuations [this can be seen from the non-linear terms, e.g., $(\delta\bb{u}\bcdot\bb{\nabla})\delta\bb{b}$, in which the contribution from $\delta\bb{u}_{\|,\lambda}$ to $\delta\bb{u}\bcdot\bb{\nabla}$, which is the one associated with the angle $\widetilde{\theta}_\lambda$, is subdominant by a factor of $k_\|/k_\perp\ll1$].

\begin{table*}
\centering
\begin{tabular}{c|cccc}

\hline
\hline

  & & & & \\
  &            & standard weak regime    & moderately weak         & asymptotically weak \\  
  & definition  & without alignment       & with alignment          & with alignment \\  
  &            & (``W0'') & (``WI'') & (``WII'') \\  
  & & & & \\
\hline
\hline
  & & & & \\
 $\widetilde{\theta}_\lambda$ & $\xi_\lambda/\ell_0$ & --- & $(\lambda/\ell_0)^{1/2}$ & const ($\ll\theta_\lambda$) \\ 
  & & & & \\
\hline
  & & & & \\
 $\theta_\lambda$ & $\lambda/\xi_\lambda$ & --- & $(\lambda/\ell_0)^{1/2}$ & $(\lambda/\ell_0)$ \\ 
  & & & & \\
\hline
  & & & & \\
 eddies &  & elongated tubes & elongated ribbons & extended sheets \\ 
 shape  &  & (``spaghetti'') & (``fettuccine'') & (``lasagne'') \\ 
  & & & & \\ 
\hline
  & & & & \\
 $\tau_{{\rm nl},\lambda}/\tau_{\rm A}$ & $\theta_\lambda^{-1}(\lambda/\ell_0)(v_{\rm A}/\delta b_\lambda)$ & $\Lambda_0^{-1/4}(\lambda/\ell_0)^{1/2}$ & $\Lambda_0^{-1/4}(\lambda/\ell_0)^{1/4}$ & $\Lambda_0^{-1/4}=\mathrm{const}$ \\ 
  & & & & \\
\hline
  & & & & \\
 $\delta b_\lambda/v_{\rm A}$ & $\Lambda_0^{1/4}\,\theta_\lambda^{-1/2}\,(\lambda/\ell_0)^{1/2}$ & $\Lambda_0^{1/4}(\lambda/l_0)^{1/2}$ & $\Lambda_0^{1/4}\left(\lambda/\ell_0\right)^{1/4}$ & $\Lambda_0^{1/4}=\mathrm{const}$ \\
  & & & & \\
\hline
  & & & & \\
 ${\cal E}_{\delta b}/(l_0 v_{\rm A}^2)$ & $\Lambda_0^{1/2}\,\theta_\lambda^{-1}\,(\lambda/\ell_0)^{2}$ & $\Lambda_0^{1/2}(\lambda/l_0)^2$ & $\Lambda_0^{1/2}\left(\lambda/\ell_0\right)^{3/2}$ & $\Lambda_0^{1/2}\left(\lambda/\ell_0 \right)$ \\
  & & & & \\
\hline
  & & & & \\
 $\lambda_{\rm CB}/{\ell_0}$ & $\tau_{{\rm nl},\lambda_{\rm CB}}\sim\tau_{\rm A}$ & $\Lambda_0^{1/2}$ & $\Lambda_0$ & --- \\
  & & & & \\
\hline
  & & & & \\
 $\lambda_*/{\ell_0}$  & $\gamma_{\lambda_*}^{\rm t}\,\tau_{{\rm nl},\lambda_*}\sim1$ & --- & $\Lambda_0^{-1/9}S_0^{-4/9}$ & $\Lambda_0^{-1/12}S_0^{-1/3}$ \\
  & & & & \\
\hline 
\hline 
\end{tabular}
\caption{Scalings pertaining to relevant quantities and critical scales when dynamic alignment is included in the weak turbulence regime (phenomenological derivation; a ``$\sim$'' relating the various quantities to their scaling is understood). A ``0'' subscript denotes quantities evaluated at the injection scale, $\lambda\sim k_\perp^{-1}$ refers to the fluctuation's wavelength perpendicular to both a scale-dependent mean magnetic field $\langle\bb{B}\rangle_\lambda$ and to the fluctuations $\delta\bb{b}_\lambda$ themselves, $\tau_{\rm A}=\ell_0/v_{\rm A}$ is the Alfv\'en (``linear'') time, and $v_{\rm A}=v_{\rm A,0}=\mathrm{const}$ is the Alfv\'en speed. We have introduced $\Lambda_0\doteq\varepsilon\ell_0/v_{\rm A}^3\sim\chi_0^2M_{{\rm A},0}^2$, where $\varepsilon\sim\varepsilon_0\sim u_0^2/\tau_0\sim\chi_0^2u_0^2/\tau_{\rm A}$ is the energy cascade (and injection) rate per unit mass, $\tau_0\sim\tau_{{\rm nl},0}^2/\tau_{\rm A}\sim\tau_{\rm A}/\chi_0^2$ is the cascade time at the outer scale, and $\chi_0=\tau_{\rm A}/\tau_{{\rm nl},0}$ and $M_{{\rm A},0}=u_0/v_{\rm A}$ are the non-linearity parameter and Alfv\'enic Mach number at injection, respectively. Finally, $\gamma_{\lambda}^{\rm t}\,\tau_{\rm A}\sim S_0^{-1/2}(\lambda/\ell_0)^{-3/2}(\delta b_\lambda/v_{\rm A})^{1/2}$ is the maximal tearing growth rate \citep{LoureiroBoldyrevPRL2017,MalletMNRAS2017}, where $S_0\doteq \ell_0 v_{\rm A}/\eta$ is the Lundquist number evaluated at the outer scale.}
\label{tab:scalings}
\end{table*}

\subsection{Dynamic alignment at weak nonlinearities}\label{subsec:theory_alignment}

The fluctuations' scaling laws are derived by assuming a constant energy flux throughout each scale of the inertial range, {\it viz.}~$\delta b_\lambda^2/\tau_\lambda\sim \varepsilon=\mathrm{const}$, and by adopting the weak-regime cascade time $\tau_\lambda\sim\tau_{{\rm nl},\lambda}^2/\tau_{\rm A}$.
Taking into account alignment in $\tau_{{\rm nl},\lambda}\sim\lambda/(\theta_\lambda \delta b_\lambda)$, this leads to $\delta b_\lambda/v_{\rm A}\sim(\varepsilon\ell_0/v_{\rm A}^3)^{1/4}\,\theta_\lambda^{-1/2}\,(\lambda/\ell_0)^{1/2}$ and ${\cal E}_{\delta b}(\lambda)\propto (\varepsilon\ell_0/v_{\rm A}^3)^{1/2}\,\theta_\lambda^{-1}\,(\lambda/\ell_0)^{2}$. For convenience, the reader can find all of these definitions in the first column of Table~\ref{tab:scalings}.

When alignment is neglected one obtains the usual weak-turbulence scalings (hereafter ``W0'') $\delta b_\lambda^{\rm(W0)}\propto\lambda^{1/2}$ and ${\cal E}_{\delta b}^{\rm(W0)}(\lambda)\propto\lambda^2$. This case does not include the  fluctuations' anisotropy in the plane perpendicular to the mean magnetic field (i.e., turbulent eddies are very elongated ``spaghetti''-like structures). It achieves CB at scale $\lambda_{\rm CB}^{\rm(W0)}/\ell_0\sim(\varepsilon\ell_0/v_{\rm A}^3)^{1/2}\approx \chi_0 M_{{\rm A},0}$, where $M_{{\rm A},0}\sim u_0/v_{\rm A}$ is the Alfv\'enic Mach number at injection (see second column of Table~\ref{tab:scalings} for a summary of these scalings). 
Next we discuss how this picture might be modified by dynamic alignment. For the sake of clarity and simplicity, we consider here only two limiting cases that may be relevant for the interpretation of our simulation results; the general case will appear in a separate publication.

Using the maximal-alignment argument by \citet{Boldyrev2006}, in which $\widetilde{\theta}_\lambda$ and $\theta_\lambda$ scale in the same way (as does $\phi_\lambda^{ub}$), one obtains $\theta_\lambda\sim\widetilde{\theta}_\lambda\propto\lambda^{1/2}$ (hereafter ``WI''). This case develops three-dimensional eddies with very elongated ``fettuccine''-like structure ($\lambda$ decreases faster than $\xi_\lambda\propto\lambda^{1/2}$, eventually attaining $\lambda\ll\xi_\lambda\ll\ell_0$). This regime is characterized by a spectrum ${\propto}k_\perp^{-3/2}$, as in \citet{Boldyrev2006}, but with $k_\|\sim\mathrm{const}$ instead of $k_\|\propto k_\perp^{1/2}$.
In this ``WI'' case, CB is reached at $\lambda_{\rm CB}^{\rm(WI)}/\ell_0\sim\varepsilon\ell_0/v_{\rm A}^3\sim \chi_0^2\,M_{{\rm A},0}^2$, i.e., at scales typically smaller than those of ``W0'' due to weaker nonlinearities induced by alignment.
All of these scalings are conveniently summarized in the third column of Table~\ref{tab:scalings}.
In this regard, we mention that a ${\approx}-3/2$ spectrum had been observed previously in high-resolution simulations of weak MHD turbulence by \citet{MeyrandJFM2015}; in these simulations, such a spectrum was found either as a large-scale range before transitioning into a smaller-scale standard weak-turbulence spectrum ${\sim} k_\perp^{-2}$, or as the fluctuations' spectrum when artificially removing $k_\|=0$ modes (which indeed do not belong to the weak regime). The authors also report the emergence of strong intermittency, which they relate to the presence of intense current sheets (and perhaps one would recognize some plasmoid-like structures as well; see their Figure 1), in the plane perpendicular to $\bb{B}_0$. Although the realization of such a field-perpendicular anisotropy would indeed require some sort of dynamic alignment, the authors did not focus on this type of analysis, so at this stage we can only mention a plausible, qualitative connection with our predicted scalings.

Considering an asymptotically weak regime with $\delta b/B_0\ll1$ (hereafter ``WII''), one can neglect the angle $\widetilde{\theta}_\lambda$ between $\langle\bb{B}\rangle_\lambda$ and $\bb{B}_0$ with respect to $\theta_\lambda$ (hence, $\phi_\lambda^{ub}\sim\theta_\lambda$). Since $\widetilde{\theta}$ is finite for finite $\delta b$, we consider $\widetilde{\theta}_\lambda\sim\xi_\lambda/\ell_0\sim\mathrm{const}\ll\theta_\lambda$, so that $\xi_\lambda\sim\mathrm{const}\sim\xi_0$ and thus $\theta_\lambda\sim\lambda/\xi_\lambda\propto\lambda$. 
This case develops eddies that shrink only in the direction defined by $\lambda$, i.e., ``lasagne''-like sheets (e.g., $\lambda\ll\xi_0<\ell_0$ for very oblique AWs; this is reminiscent of our $\chi_0\sim0.1$ simulation). This regime is characterized by scale-invariant fluctuations $\delta b_\lambda\sim\const$, which thus produce a spectrum ${\propto}k_\perp^{-1}$, and by the fact that the cascade never reaches CB (because alignment depletes the nonlinearities so that $\tau_{{\rm nl},\lambda}\sim\mathrm{const}$). 
These scalings are repeated in the last column of Table~\ref{tab:scalings}.
In this regard, it is worth mentioning that (rapid) scale-dependent alignment between $\delta\bb{u}_\perp$ and $\delta\bb{b}_\perp$ fluctuations (or, anti-alignment between $\delta\bb{z}_\lambda^+$ and $\delta\bb{z}_\lambda^-$) has been reported to occur in the $-1$ spectral range of solar-wind turbulence by \citet{WicksPRL2013}. This ensemble of aligning fluctuations was measured to constitute the majority of the fluctuations' population and to be the one responsible for the resulting $-1$ spectrum (structure functions show scale-independent behavior of $\delta b_\perp$ in that range; their Figure 1); they were interpreted as ``non-turbulent'' fluctuations, i.e., belonging to non-interacting (or, weakly interacting) counter-propagating AWs. Despite the fact that here we do not take into account the effect of imbalance or residual energy, one may relate the fluctuations' behavior in \citet{WicksPRL2013} to the basic ideas underlying our ``WII'' case. In fact, in order to have dynamic alignment, such a population of counter-propagating AWs have to be shearing one another---and thus have a small, but finite, amount of nonlinear interactions (i.e., to be in the asymptotically weak $\chi\ll1$ regime discussed above).

\subsection{Dynamically aligned, weak turbulence meets reconnection}\label{subsec:theory_reconnection_scale}

Given the above scalings, cascading fluctuations should develop an anisotropy perpendicular to $\langle\bb{B}\rangle_\lambda$ that increases significantly faster than the one associated with a strong cascade~\citep[$\xi/\lambda\propto\lambda^{-1}$ for ``WI'' and ${\propto}\lambda^{-1/2}$ for ``WII'', instead of ${\propto}\lambda^{-1/4}$ in][]{Boldyrev2006}. At the same time, turbulent eddies at a given scale live longer for weaker nonlinearities, leaving more time for tearing instability to grow.
Thus, a transition to tearing-mediated turbulence should occur at larger scales when starting from a weakly nonlinear regime. 

The critical scale $\lambda_*$ at which tearing can grow on top of turbulent eddies is determined by requiring that the tearing growth timescale is comparable to the eddy lifetime, $\gamma_{\lambda_*}^{\rm t}\tau_{{\rm nl},\lambda_*}\sim1$. 
Following \citet{LoureiroBoldyrevPRL2017} and \citet{MalletMNRAS2017}, we adopt $\gamma_{\lambda}^{\rm t}\,\tau_{\rm A}\sim S_0^{-1/2}(\lambda/\ell_0)^{-3/2}(\delta b_\lambda/v_{\rm A})^{1/2}$ for the maximal tearing growth rate, where $S_0\doteq \ell_0 v_{\rm A}/\eta$ is the Lundquist number evaluated at the outer scale. 
The transition scales in the ``WI'' and ``WII'' limits are summarized in Table~\ref{tab:scalings}.

The ``WI'' case can either transition to the  ``\citet{Boldyrev2006}-type'' of strong turbulence or to a tearing-mediated cascade: since $\lambda_*^{\rm(WI)}/\lambda_{\rm CB}^{\rm(WI)}\sim \chi_0^{-20/9}M_{{\rm A},0}^{-20/9}S_0^{-4/9}$, this means that tearing-mediated turbulence will prevail over the critically balanced cascade \`a la \citet{Boldyrev2006} when $\chi_0<M_{{\rm A},0}^{-1}S_0^{-1/5}$. In this case, one requires only that $M_{{\rm A},0}\sim0.1$ and $S_0\sim10^{5}$ for the transition to the usual critically balanced cascade to be replaced by a transition to a tearing-mediated range for any $\chi_0<1$. 
On the other hand, tearing completely replaces the usual CB transition in case ``WII''.  
For instance, adopting a fixed Lundquist number $S_0$ across all regimes, one finds that $\lambda_*^{\rm(WI)}/\lambda_*^{\rm(CB)}\sim(\chi_0 M_{{\rm A},0})^{-2/9}S_0^{8/63}$ and $\lambda_*^{\rm(WII)}/\lambda_*^{\rm(CB)}\sim(\chi_0 M_{{\rm A},0})^{-1/6}S_0^{5/21}$, where $\lambda_*^{\rm(CB)}/\ell_0\sim S_0^{-4/7}$ is the predicted transition scale in the strong, critically balanced regime (for $\chi_0\sim1$ and $M_{{\rm A},0}\sim 1$)~\citep{LoureiroBoldyrevPRL2017,MalletMNRAS2017,BoldyrevLoureiroAPJ2017}. More specifically, using the parameters of our $\chi_0\sim0.5$ simulation for case ``WI'', we predict a transition scale that is ${\approx}6$ times larger than the corresponding scale in the strong regime. Analogously, employing the parameters of the $\chi_0\sim0.1$ simulation for case ``WII'', we find a transition scale that would be ${\sim}60$ times larger than the one predicted following a cascade \`a la \citet{Boldyrev2006}.

\subsection{Conjecture of tearing-driven CB}\label{subsec:theory_reconnection_CB}

At this point, one may be tempted to derive the scalings for the tearing-mediated range in the weak regime by substituting $\tau_{{\rm nl},\lambda}$ with $\tau_\lambda^{\rm t}\sim1/\gamma_\lambda^{\rm t}$ in the cascade time\footnote{Incidentally, this would lead to $\delta b_\lambda^{\rm(W,t)}/v_{\rm A}\sim(\varepsilon\ell_0/v_{\rm A}^3)^{1/3}S_0^{1/3}(\lambda/\ell_0)$, so that the spectrum would be ${\cal E}_{\delta b}^{\rm(W,t)}\propto\lambda^3\sim k_\perp^{-3}$ and the alignment angle at $\lambda<\lambda_*$ would increase (i.e., fluctuations would misalign) as $\theta_\lambda^{\rm(W,t)}\propto\lambda^{-1}\sim k_\perp$.}, so that $\tau_\lambda\sim(\tau_\lambda^{\rm t})^2\,\tau_{\rm A}^{-1}$.
However, a main feature of the weak regime, namely that $\ell_\lambda=\ell_0=\mathrm{const}$, cannot hold if the cascade is mediated by tearing. This is because tearing will produce reconnecting magnetic islands and thereby generate smaller scales in the magnetic-field fluctuations, both in the perpendicular direction ($\lambda$ and $\xi_\lambda$) and the parallel direction ($\ell_\lambda$). 
How would $\ell_\lambda$ change, then? Since $\tau_\lambda^{\rm t}$ is now the timescale over which  $\delta b_\lambda$ fluctuations are generated at $\lambda<\lambda_*$, it is reasonable to consider that timescale to be the actual transfer time, {\em viz.}, $\tau_\lambda\sim\tau_\lambda^{\rm t}$. Therefore, because of the condition $\tau_\lambda^{\rm t}\sim\tau_{{\rm nl},\lambda}$ and the fact that $\tau_\lambda\sim \tau_{{\rm nl},\lambda}^2/\tau_{\rm A}$ holds up to scale $\lambda\sim\lambda_*$, it follows that $\tau_\lambda^{\rm t}\sim\tau_{{\rm A},\lambda}$ (note that $\tau_{\rm A}$ is not scale-independent anymore below $\lambda_*$). 
This argument can explain the reduced number of AW-packet interactions required to achieve a fully developed turbulent state in our $\chi_0\sim0.5$ and $\chi_0\sim0.1$ simulations, {\em viz.}, $N_{\rm int}^*\propto\chi_0^{-1}$ (Figure~\ref{fig:Jrms_compare_hst}, left-panel inset).
This indicates that, at scales $\lambda<\lambda_*$, CB should be expected to hold. We therefore conjecture that tearing drives the cascade towards CB and to the usual $-11/5$ spectrum of tearing-mediated turbulence.

\section{Discussion and Conclusions}\label{sec:discussion-conclusion}

Using 3D gyro-fluid simulations, we have investigated how the turbulent dynamics arising from collisions of counter-propagating AW packets with different large-scale nonlinearity parameter $\chi_0$ is modified by tearing instability.

For strong initial nonlinearities ($\chi_0\sim1$), we observe a regime consistent with dynamically aligned, critically balanced MHD turbulence~\citep{Boldyrev2006}, i.e., fluctuations align accordingly to $\sin\theta_{k_\perp}\sim k_\perp^{-1/4}$, resulting in a $k_\perp^{-3/2}$ spectrum with $k_\|\propto k_\perp^{1/2}$ spectral anisotropy. Tearing does not appear to modify the cascade, consistent with theoretical expectations given the Lundquist numbers we are able to afford in our numerical simulations.

As the initial nonlinearities are lowered ($\chi_0<1$), however, a spectral break marking the transition between large-scale weak turbulence and small-scale tearing-mediated turbulence appears. 
The presence of a tearing-mediated range for small $\chi_0$ implies that dynamic alignment occurs also at weak nonlinearities.
In particular, for these cases the alignment angle shows a stronger scale dependence than found in the critically balanced regime, namely $\sin\theta_{k_\perp}\sim k_\perp^{-1/2}$ at $\chi_0\sim0.5$, and $\sin\theta_{k_\perp}\sim k_\perp^{-1}$ at $\chi_0\sim0.1$: this, combined with the increased lifetime of turbulent eddies at small $\chi_0$, allows tearing to onset and mediate the cascade at scales larger than those predicted for a strong MHD cascade. 
Dynamic alignment in the weak regime also determines a modification to the large-scale spectrum, roughly scaling as $k_\perp^{-3/2}$ for $\chi_0\sim0.5$ and as $k_\perp^{-1}$ for $\chi_0\sim0.1$.

Regardless of the large-scale nonlinearity parameter, the emerging tearing-mediated range is consistent with the predicted $k_\perp^{-11/5}$ spectrum and a scale-dependent (mis)alignment of the fluctuations following something close to $\sin\theta_{k_\perp}\sim k_\perp^{4/5}$~\citep{MalletMNRAS2017,BoldyrevLoureiroAPJ2017,ComissoAPJ2018}. 
These scalings, together with the fact that in our simulations the number of AW-packet interactions necessary to achieve a fully developed turbulent state for these low-$\chi_0$ regimes is reduced with respect to the weak-turbulence expectation ({\em viz.} $\propto\chi_0^{-1}$ instead of $\propto\chi_0^{-2}$), support our conjecture of a ``tearing-induced'' transition to CB.

A phenomenological theory of dynamically aligned turbulence at weak nonlinearities that can explain these spectra and the transition to the tearing-mediated regime is provided.
In particular, it is shown that, depending on the nonlinearity parameter at injection and on the large-scale Alfv\'enic-Mach and Lundquist numbers, the transition to tearing-mediated turbulence may compete (if not completely supplant) the usual transition to CB; and that such a transition scale at small nonlinearities can be larger than the one implied by a critically balanced MHD cascade by several orders of magnitude, if the Lundquist number of the system is large enough~\citep[cf.][]{MalletMNRAS2017,BoldyrevLoureiroAPJ2017,ComissoAPJ2018}.
We expect such a shift of the transition scale $\lambda_*$ to scales larger than those implied by a strong MHD cascade to be a general consequence of the fact that dynamic alignment occurs also in the weak regime, regardless of the precise physics of tearing (i.e., resistive or collisionless); the precise scaling of such a transition scale, on the other hand, will clearly depend upon the micro-physics of tearing~\citep[e.g.,][]{LoureiroBoldyrevAPJ2017,MalletJPP2017}.

Our results suggest a more complex scenario than the simplistic picture of weak-to-strong transition in Alfv\'enic turbulence and shed new light on the existence of different large-scale regimes that coexist with tearing-mediated turbulence. This may have significant implications for small-scale dissipation and turbulent heating in space and astrophysical plasmas. 
Moreover, depending on the Lundquist number, a dynamically aligned weak cascade will undergo a transition to tearing-mediated turbulence at scales larger than the scales at which a standard weak cascade would meet the usual CB condition. Because this implies that a cascade in $k_\|$ is realized earlier in $k$ (and with larger fluctuation amplitudes), our new scalings may have significant implications on the scattering efficiency of cosmic rays in astrophysical environments in which Alfv\'enic turbulence is injected with small nonlinearities and/or at small Alfv\'enic-Mach numbers~\citep[e.g.,][]{ChandranPRL2000,YanLazarianPRL2002,YanLazarianAPJ2008,FornieriMNRAS2021,KempskiQuataertMNRAS2022}.

Finally, our results and the basic ideas underlying our new scalings can be viewed in connection with {\em in-situ} measurements of solar-wind turbulence. 
For instance, a (rapid) scale-dependent alignment between $\delta\bb{u}_\perp$ and $\delta\bb{b}_\perp$ fluctuations (or, anti-alignment between $\delta\bb{z}_\lambda^+$ and $\delta\bb{z}_\lambda^-$) has been reported to occur in the large-scale $-1$ range of solar-wind turbulence by \citet{WicksPRL2013}. In particular, it was shown that such an ensemble of aligning fluctuations constitutes the majority of the fluctuations' population, and that they are responsible for the resulting $-1$ spectrum ({\em viz.}, structure functions reveal a scale-independent behavior of $\delta b_\perp$ in that range); these fluctuations were interpreted as ``non-turbulent'' fluctuations belonging to quasi-non-interacting, counter-propagating AWs. Since a finite, whatever small, amount of nonlinear interactions is required to occur for counter-propagating AWs to be shearing one another and induce dynamic alignment, we suggest that this may be the case for the aligning population observed by \citet{WicksPRL2013}, thus potentially pertaining to an asymptotically weak ($\chi\ll1$) regime as discussed in our scalings (\S\ref{subsec:theory_alignment}, case ``WII''). 
Another intriguing piece of {\em in-situ} measurement is the one recently taken by {\em Parker Solar Probe}  within the magnetically dominated corona~\citep{KasperPRL2021}. Among other features, the magnetic-field spectrum in that region exhibits a transition between a $-3/2$ range and a steeper $\approx-2.2$ slope occurring at scales (frequencies) much larger (smaller) than the ion characteristic scales (frequencies), which may be a hint of a potential large-scale, tearing-mediated range. While further studies are definitely needed to investigate the fluctuations' properties across this transition (e.g., estimated strength of nonlinearities, spectral anisotropy, etc.) in order to understand what type of transition we are observing, our theory in the moderately weak regime (\S\ref{subsec:theory_alignment}, case ``WI'') provides an alternative scenario to  interpret the measurements by \citet{KasperPRL2021}.

While the underlying processes highlighted by the above {\em in-situ} spacecraft measurements may be the same on which our scalings are founded (namely, dynamic alignment in the weak regime, followed by a large-scale transition to tearing-mediated turbulence), we caution as a final remark that these connections are purely conceptual, as our theory does not take into account imbalance or residual energy. With these being outside the scope of the current work, a more detailed theory that also includes these effects will be explored in a following paper.

\acknowledgements 
The authors warmly acknowledge constructive advice from the anonymous referee, as well as productive comments by and discussions with Alexander Schekochihin following our posting of the original manuscript to arXiv. 
SSC also acknowledges useful discussions with Alexandre Lazarian, after reporting the results presented in this manuscript at the 44th COSPAR Scientific Assembly in Athens, Greece.
We gratefully acknowledge access to the Stellar cluster at the PICSciE-OIT TIGRESS High Performance Computing Center and Visualization Laboratory at Princeton University, where the main simulations and analysis were performed. Computations were also performed on the ``Mesocentre SIGAMM'' machine, hosted by Observatoire de la Côte d’Azur.
The authors warmly thank the organizers of the ``Waves C\^ote d'Azur'' conference in Nice (June 2019), where the first discussions leading to this work took place. MWK thanks the Institut de Plan\'etologie et d'Astrophysique de Grenoble (IPAG) for its hospitality and visitor support while this work was completed.

\appendix

\section{A.~Numerical resolution, Lundquist number, and dissipation operators}\label{app:numerical_tests}

In this Appendix, we summarize the outcome of various numerical tests that have been performed in preparation for the production runs. These tests focused on (i) the effectiveness of the numerical dissipation, (ii) the ability to identify clearly a tearing-mediated range, and (iii) the effect of employing different dissipation operators.

\begin{figure*}[!b]
\center%
\includegraphics[width=0.5\textwidth]{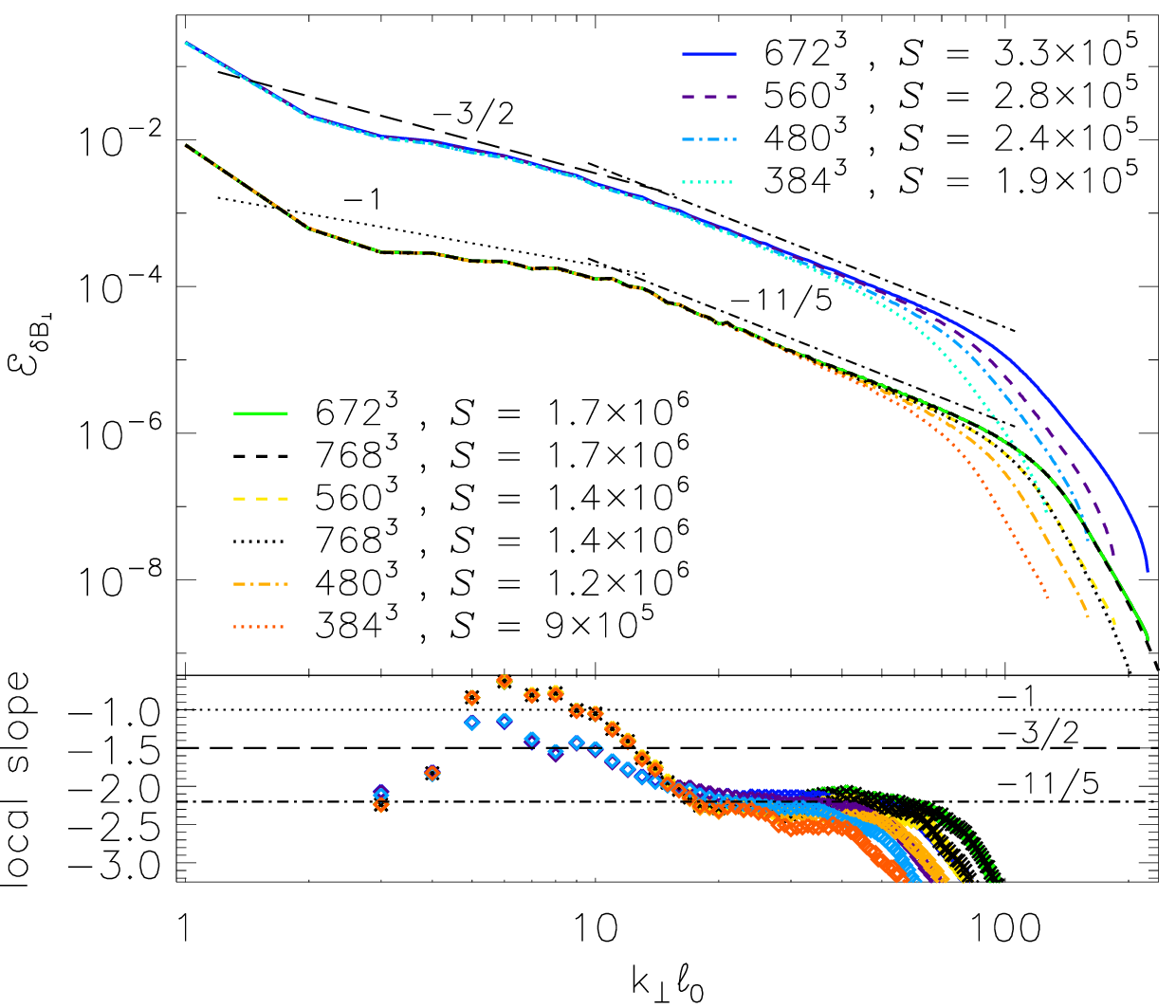}%
\includegraphics[width=0.5\textwidth]{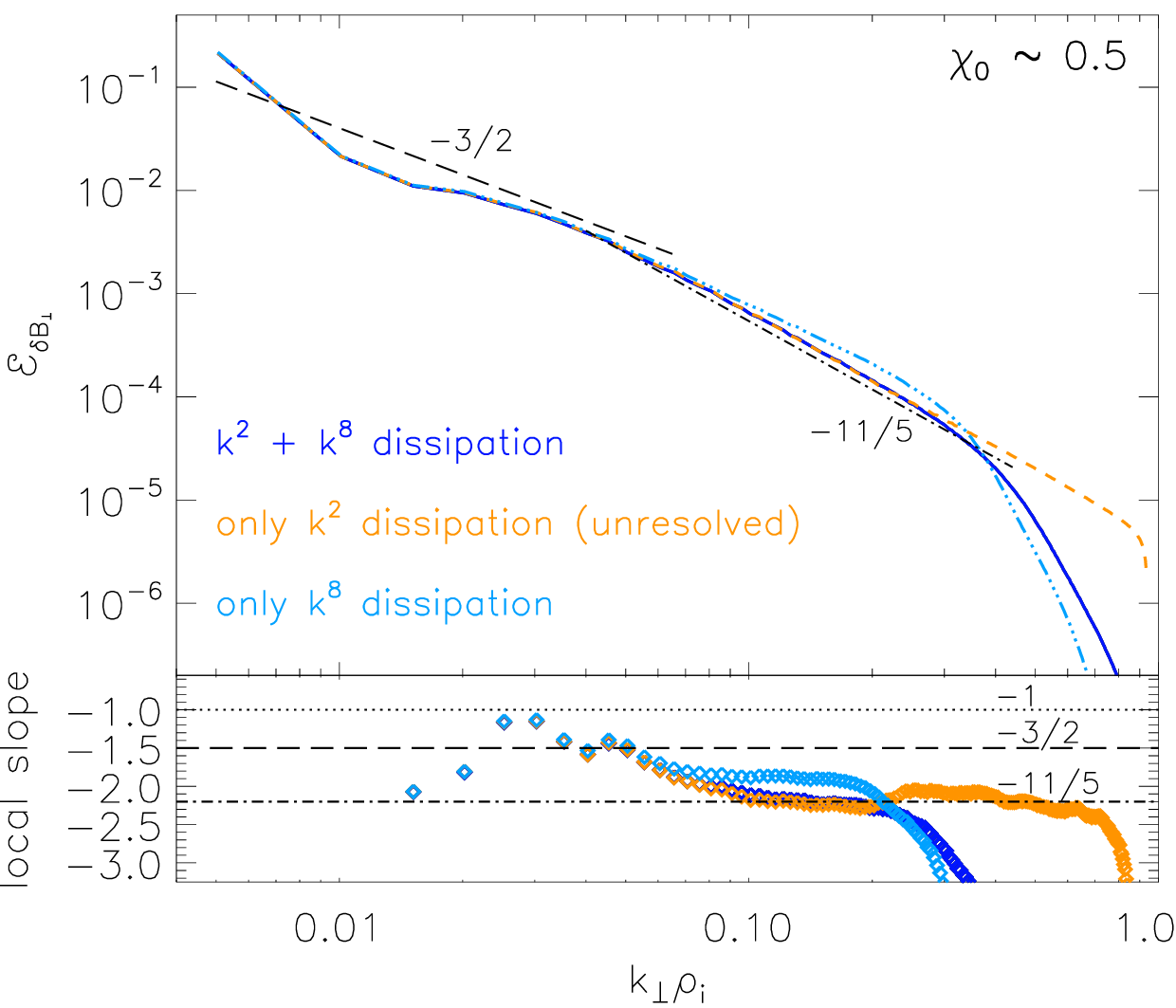}%
\caption{Left: Time-averaged energy spectrum of $\delta B_\perp$ fluctuations and its local slope versus $k_\perp\ell_0$ for different Lundquist numbers and/or resolutions, at $\chi_0\sim 0.5$ (upper spectra, in blue shades and different line styles) and at $\chi_0\sim0.1$ (lower spectra, green/yellow/orange/black colors and different line styles).  The Lundquist number $S=L_0 v_{\rm A}/\eta$ has been varied by changing $L_0$ at fixed resolution. Two examples of different resolutions at fixed $\chi_0\sim0.1$ exhibit the same behavior ($L_0\simeq1244\rho_{\rm i}$ and $S\simeq1.4\times10^6$, yellow-dashed and black-dotted lines; $L_0\simeq1493\rho_{\rm i}$ and $S\simeq1.7\times10^6$, green-solid and black-dashed lines), demonstrating numerical convergence for the chosen dissipation parameters. Right: Time-averaged $\delta B_\perp$ spectrum and its local slope versus $k_\perp\rho_{\rm i}$ at $\chi_0\sim0.5$ for different combinations of dissipation operators, viz., only Laplacian dissipation ($\propto k^2$, orange dashed line), only 8th-order dissipation ($\propto k^8$, light-blue dash-three-dotted line), and a combination of both Laplacian and 8th-order dissipation (blue solid line; this corresponds to the purple dashed line in the left panel, i.e., $S\simeq2.8\times10^5$). This shows that, when a combination of the two operators is employed, our choice of dissipation parameters is such that the tearing-mediated range ${\propto} k_\perp^{-11/5}$ is due to the usual (i.e., Laplacian) resistivity.}
\label{fig:convergence_tests}  
\end{figure*}

Point~(i) has been addressed by increasing the resolution while fixing the large-scale properties and the dissipation parameters. The results of this convergence test are shown in the left panel of Figure~\ref{fig:convergence_tests}, in which the time-averaged spectrum of $\delta B_\perp$ (and its local slope) versus $k_\perp\ell_0$ is reported for the $\chi_0\sim0.1$ regime and for two large-scale Lundquist-number cases, $S\simeq1.4\times10^6$ (yellow-dashed and black-dotted lines) and $S\simeq1.7\times10^6$ (green-solid and black-dashed lines), at different small-scale resolutions. The overlap of the spectra and of their local slopes at increased resolution shows the effectiveness of the dissipation parameters employed (hereafter, referred to as ``optimal'').

Point (ii) has been addressed by keeping the small-scale resolution and (optimal) dissipation parameters fixed, while the Lundquist number $S=L_0 v_{\rm A}/\eta$ has been varied by changing the injection scale $L_0$. A summary of this study is reported in the left panel of Figure~\ref{fig:convergence_tests}, in which spectra of $\delta B_\perp$ (and their local slopes) versus $k_\perp\ell_0$ are shown for different Lundquist numbers at $\chi_0\sim 0.5$ (upper spectra, in blue shades and different line styles) and at $\chi_0\sim0.1$ (lower spectra, green/yellow/orange/black colors and different line styles). Although the existence of a $-11/5$ range seems to be visible already at the lowest Lundquist numbers, a reliable and fairly extended tearing-mediated spectrum is obviously achieved only at the largest separation of scales (i.e., larger $L_0$ at fixed dissipation scales, corresponding to larger $S$). In this context, the values $S\simeq1.7\times10^6$ and $S\simeq3.3\times10^5$ for the $\chi_0\sim0.1$ and $\chi_0\sim0.5$ regimes, respectively, were considered to provide a satisfactory result.

Finally, the impact of the dissipation order has been explored by varying the operators employed and/or by taking a combination of different orders. The outcome is summarized in the right panel of Figure~\ref{fig:convergence_tests}, which shows spectra of $\delta B_\perp$ and their local slopes versus $k_\perp\rho_{\rm i}$ in the $\chi_0\sim0.5$ regime when using: (a) only a Laplacian operator ($\propto k^2$, orange dashed line; note that to obtain a $-11/5$ range for this case, the dissipation level is insufficient and thus energy accumulates at the smallest scales of the system--and so the simulation is considered to be ``unresolved''); (b) only an 8th-order operator ($\propto k^8$, light-blue dash-three-dotted line); or (c) a combination of Laplacian and 8th-order operators (blue solid line; this simulation corresponds to the purple dashed line in the left panel of Figure~\ref{fig:convergence_tests}, i.e., $S\simeq2.8\times10^5$ on a $560^3$ grid). Although the fluctuations' spectrum for the case with only Laplacian dissipation clearly shows a slight rise of the spectral slope at $k_\perp\rho_{\rm i}>0.2$, it overlaps in the range $k_\perp\rho_{\rm i}\lesssim0.2$ with the spectrum obtained from the (well-resolved) simulation employing both Laplacian and 8th-order dissipation operators. We are therefore confident that, in the latter case, the break scale $k_\perp^*\rho_{\rm i}\approx0.07$ and the $-11/5$ slope in the range $0.08\lesssim k_\perp\rho_{\rm i}\lesssim0.2$ are due to the usual resistive reconnection (while the additional hyper-resistivity simply completes the energy dissipation at the smallest scales of the system). On the other hand, there is no clear signature of a tearing-mediated $-11/5$ range in the spectrum obtained from the simulation employing only an 8th-order dissipation operator; although an apparent break at $k_\perp\rho_{i}\sim0.06$ followed by a $\approx-1.85$ slope in the range $0.07\lesssim k_\perp\rho_{\rm i}\lesssim0.15$ seem to be present, these may be due to other effects rather than (hyper-resistive) reconnection~\citep[which would instead exhibit a $-19/7\approx-2.7$ slope, according to predictions by][]{BoldyrevLoureiroAPJ2017}. The reason for this slope in the purely hyper-resistive case is not clear at this stage, and will require further investigation.
\\


\end{document}